\documentclass[11pt,a4paper]{article}
\pdfoutput=1

\usepackage{jcappub}

\usepackage{amssymb,amsmath,amsthm}

\usepackage{upgreek}
\usepackage{mathtools}

\usepackage{dsfont}
\usepackage{slashed}
\usepackage{cancel}
\usepackage{fullpage}
\usepackage{graphicx}

\usepackage{mathrsfs}

\usepackage[font=small,labelfont=bf]{caption} 
\usepackage{float}

\usepackage{braket}
\usepackage{color}
\usepackage{siunitx}
\usepackage{xcolor}
\usepackage{cancel}
\usepackage{bm}

\usepackage{epstopdf}

\usepackage{natbib}

\usepackage{hyperref}
\hypersetup{
     colorlinks   = true,
     citecolor    =  red
}
\usepackage{cleveref}

\usepackage{amsmath,amssymb}

\usepackage{mathtools}
\usepackage{braket}
\usepackage{siunitx}
\usepackage{empheq}

\usepackage{tikz}
\usetikzlibrary{decorations.pathmorphing}
\usetikzlibrary{calc}

\usepackage{subcaption}

\newcommand\mi{\mathrm{i}} 
\newcommand\me{\mathrm{e}} 

\usepackage{upgreek}

\newcommand\pp{\uppi}

\newcommand{\dif}{\mathrm{d}}

\DeclareMathOperator{\arccoth}{arccoth}

\DeclareMathOperator{\BesselJ}{J}

\DeclareMathOperator{\BesselF}{F}
\DeclareMathOperator{\BesselG}{G}

\DeclareMathOperator{\sech}{sech}
\DeclareMathOperator{\csch}{csch}

\newcommand{\rbr}[1]{{\left(#1\right)}}
\newcommand{\sbr}[1]{{\left[#1\right]}}

\newcommand{\vbr}[1]{{\left|#1\right|}}

\newcommand*\abs[1]{\left|#1\right|}

\newcommand{\rfun}[2]{{#1}\mathopen{}\left(#2\right)\mathclose{}}
\newcommand{\sfun}[2]{{#1}\mathopen{}\left[#2\right]\mathclose{}}

\def\beq{\begin{equation}}
\def\eeq{\end{equation}}

\def\a{\alpha}
\def\kap{\varkappa}

\makeatletter
\newcommand*{\rom}[1]{\expandafter\@slowromancap\romannumeral #1@}
\makeatother

\title{Integrable Minisuperspace Models with Liouville
Field: Energy Density Self-Adjointness and Semiclassical Wave Packets}

\author[a,d]{Alexander A.\ Andrianov,}
\author[b]{Chen Lan,}
\author[a]{Oleg O.\ Novikov}
\author[c]{and \\ Yi-Fan Wang}

\affiliation[a]{Saint Petersburg State University, \\
7/9 Universitetskaya nab., St. Petersburg 199034, Russia}
\affiliation[b]{ELI-ALPS Research Institute, \\
Budapesti \'ut 5, H-67228 Szeged, Hungary}
\affiliation[c]{Institut f\"ur Theoretische Physik, Universit\"at zu K\"oln, \\
Z\"ulpicher Stra\ss e 77, D-50937 K\"oln, Germany}
\affiliation[d]{Institut de Ci\`encies del Cosmos, Universitat de Barcelona, \\
Mart\'i i Franqu\`es 1, E-08028 Barcelona, Spain}

\emailAdd{a.andrianov@spbu.ru}
\emailAdd{stlanchen@yandex.ru}
\emailAdd{o.novikov@spbu.ru}
\emailAdd{yfwang@thp.uni-koeln.de}

\abstract{
The homogeneous cosmological models with a Liouville scalar field
are investigated in classical and quantum context of Wheeler-DeWitt
geometrodynamics.
In the quantum case of quintessence field with potential unbounded from below
and phantom field, the energy density operators  are not essentially 
self-adjoint and self-adjoint extensions contain ambiguities. Therefore the same 
classical actions correspond to a family of distinct quantum models. For the phantom field 
the energy spectrum happens to be discrete.
The 
probability conservation and appropriate classical limit can be achieved with a certain restriction of the functional class. The appropriately localized wave packets are studied numerically using the 
Schr\"odinger's norm and a conserved Mostafazadeh's norm introduced from 
techniques of pseudo-Hermitian quantum mechanics. These norms give a similar 
packet evolution that is confronted with analytical classical solutions.
}

\keywords{quantum geometrodynamics, self-adjoint extension, integrability}

\makeatletter
\def\@fpheader{\relax}
\makeatother

\begin{document}

\maketitle



\section{Introduction}
\label{sec:introduction}
Cosmological models with scalar fields have drawn a lot of attention in the
last decades because of investigations on cosmological inflation
\cite{linde2008} and dark energy \cite{copeland2006}, but few of them can be
exactly integrated. A universe driven by scalar fields with an exponential
potential is dubbed \emph{Liouville cosmology}, which is one of the
well-studied integrable models in cosmology. The power-law expansion of
particular solutions and its applications are investigated in e.g.\
\cite{halliwell1987,gorini2004,dabrowski2006}. The general classical solutions
have been discussed in detail under various gauge conditions in e.g.\
\cite{andrianov2011,andrianov2012}. The correspondence between Jordan and
Einstein frame is studied in \cite{kamenshchik2014,kamenshchik2015a,%
kamenshchik2015b,kamenshchik2016,kamenshchik2017a,kamenshchik2017b}, wherein
the Liouville field in the Einstein frame is related to the power-law potential
in Jordan frame through a conformal transformation combining with a parameter
transformation of scalar field. The exactly solvable models with several Liouville scalar fields
were developed in \cite{andrianov2015,andrianov2016}. The appearance of the Lioville
cosmologies from higher-dimensional theories, in particular superstring theories and M-theory
 was studied in \cite{Ohta2003,Ohta2005}.

General relativity is a  theory with constraints, the corresponding Hamiltonian is
zero \cite{barvinsky1993,fulop1999,kiefer2012,barvinsky2014}.
The reason for the vanishing Hamiltonian is the presence of a non-dynamical
symmetry, namely diffeomorphism invariance; in other words, the gravitational
theory contains redundant degrees of freedom. In the minisuperspace
approximation, the redundancy appears in the form of the lapse function
$N(t)$. Therefore, to solve the dynamics of the model, it is necessary to
introduce a specific gauge condition to eliminate $N(t)$ \cite{andrianov2011,andrianov2015}.
Traditionally, the lapse function is set to unity, such that the universe evolves in cosmic time
\cite{gorbunov2016}. However one  could eliminate $N(t)$ and avoid an explicit time parametrization to obtain exact solutions of
Einstein's equation. This fits well the Wheeler--DeWitt
quantum cosmology which does not involve time.

The cosmological models driven by a scalar field with a constant potential may serve as examples of the latter approach \cite{dabrowski2006,barvinsky2014}. In these models, the scalar field
is a cyclic coordinate, hence the conjugate momentum is integral of motion, and the
conservation law can be applied to eliminate the lapse function $N(t)$, such
that the modified Friedman equation contains only minisuperspace variables.
Inspired by this, we introduce a similar integral of motion in Liouville
cosmology of homogeneous and isotropic
models  \citep{lanchen2016}, in order to eliminate the redundant degrees of freedom. With the help
of this integral of motion, the classical Friedman equation reduces to a
time-independent nonlinear equation, the solution of which can be derived
explicitly and describes the trajectory in minisuperspace.
This method can also be directly extended to higher dimensional \cite{garcia2007,letelier2010} and anisotropic models, such as Bianchi-\rom{1} cosmology considered in \cite{kamenshchik2017a}.

The physical meaning of the formal Wheeler--DeWitt equation and
its correspondence with the classical theory can be derived in three steps.
The first one is the selection of the space of \emph{physical} wave functions,
usually by endowing proper boundary conditions. In
traditional quantum mechanics, crucial properties of the theory depend on the
boundary conditions for wave functions, such as the Hermiticity of observables
\cite{essin2006}, the orthogonality of wave functions (e.g.\
\cite{araujo2004,essin2006}) and the conservation of
probability, to name a few. A similar situation holds in quantum cosmology
\cite{bouhmadi2009,barvinsky2014}, in which proper boundary conditions have to
be specified, such that the solutions of the Wheeler--DeWitt equation, which are
not square-integrable, are eliminated from the space of physical wave
functions. In this paper we address an important issue
encountered at this step. The Hamiltonian operator naively constructed by the canonical quantization in some cosmological models, which are interesting from the phenomenological point of view, including phantom field,
happens to be not essentially self-adjoint and its self-adjoint extension is not unique \cite{bonneau2001,Gitman2012,hall2013}. Namely while the clasical action fixes up to the usual ordering ambiguities how the Hamiltonian acts on the localized wavefunctions the evolution over finite amounts of time depends on its behaviour at infinity where extra ambiguity arises. Hence one classical action correspond to a family of distinct quantum models with different quantum evolutions. The cosmological models with similar self-adjointness issues
were considered in \cite{Almeida2015,Gryb2018}.

The second step is to define an inner product on the physical space that would
give the conserved probability distribution in quantum cosmology. Since the
Wheeler--DeWitt equation is of Klein--Gordon type, the `probability density'
defined by the so-called \emph{Klein--Gordon norm} is not guaranteed to be
positive. While one may restrict consideration to the WKB wavepackets the question arises how to interpret the wavefunction of the universe beyond the WKB region. A resolution of this problem may be provided within the pseudo-Hermitian theory by introducing the Mostafazadeh's norm \cite{mostafazadeh2002,mostafazadeh2006,mostafazadeh2010} . While we do not treat this norm as the only possible way to tackle the probability problem it may be considered as an useful tool to study the quantum cosmology as a fully consistent quantum theory within restrictions of the minisuperspace approximation.

Finally one has to attribute a proper energy distribution to
construct a wave packet \cite{kiefer1990,kiefer1988,kiefer1995}. For a given initial coordinate distribution of wave packet in
minisuperspace, the energy distribution can be calculated, which however is
not easy to realize in practice. A common compromise is to choose a Gaussian energy
distribution. Then in correspondence with classical theory  the probability
distribution of the established wave packet should  `centre' at the classical
path and follow it as closely as possible apart from turning points.

This paper is organized as follows. In Sec.~\eqref{sec:unstable} we briefly elucidate the problem of the quantum particle
 in the unstable potential $V=-\me^{2x}$ and the ambiguity of
self-adjoint extension of the Hamiltonian operator. In Sec.~\eqref{sec:classical} an integral of
motion is introduced for three types of Liouville cosmological models and
explicit classical solutions are given in terms of
minisuperspace variables. Sec.~\eqref{sec:quantization} introduces the corresponding
canonical quantum cosmology and there the physical state space is constructed.  As a verification of the results, in
Sec.~\eqref{sec:limit} the limit of potential parameter $\lambda$ tending to zero is
considered.  Sec.~\eqref{sec:semiclassical} is devoted to the classical-quantum
correspondence, in which the wave packets are implemented and the probability distributions are plotted for
two kinds of norms.
The conclusions Sec.~\eqref{sec:conclution} contain some comments on further extensions and applications of the approach adopted in this paper.

\section{Quantum mechanics of a particle in a negative Liouville potential}
\label{sec:unstable}

To explain the issues that will arise in the quantum cosmological models of interest let us consider the one-dimensional motion of a non-relativistic particle in a
Liouville potential which is \emph{unbounded from below}, described by the
Hamiltonian
\begin{equation}
H=p^2-\me^{2x}.\label{HamilApp}
\end{equation}
This is the special case of the unstable Morse potential considered in detail in \cite[Ch. 8.5]{Gitman2012}.
The corresponding time-independent Schr\"odinger equation is
\begin{equation}
\hat{H}\psi \coloneqq
\left(-\partial_x^2 -\me^{2x} \right) \psi = E \psi,
\label{eq:negliouvschroed}
\end{equation}

For the positive energies $E>0$ the solutions are, 
\begin{equation}
\psi_k = c_1 \rfun{\BesselF_{\mi k}}{\me^x}
	+c_2 \rfun{\BesselG_{\mi k}}{\me^x},\quad E=k^2
\end{equation}
where,
\begin{align}
\rfun{\BesselF_{\nu}}{z}&=\frac{1}{2}\,\rfun{\sec}{\frac{\nu \pp}{2}}
\left[{\BesselJ}_{\nu} (z) +{\BesselJ}_{- \nu} (z)\right],  \\
\rfun{\BesselG_{\nu}}{z} &= \frac{1}{2}\,\rfun{\csc}{\frac{\nu\pp}{2}}
\left[{\BesselJ}_{ \nu} (z) -{\BesselJ}_{- \nu} (z)\right]
\end{align}
with
\begin{equation}
\rfun{\BesselF_{\nu}}{z}={\BesselF}_{- \nu}(z),\quad
\rfun{\BesselG_{\nu}}{z} = \rfun{\BesselG_{-\nu}}{z}
\end{equation}
are defined according to \cite{dunster1990}. They have undamped oscillatory behavior as $x\rightarrow -\infty$,
\begin{align}
\rfun{\BesselF_{\mi k}}{\me^{x}}\simeq \sqrt{\frac{2\tanh\frac{\pp k}{2}}{\pp k}}\cos(kx-\delta_k)+O(\me^{2x}),\\
\rfun{\BesselG_{\mi k}}{\me^{x}}\simeq \sqrt{\frac{2\coth\frac{\pp k}{2}}{\pp k}}\sin(kx-\delta_k)+O(\me^{2x}),
\end{align}
and oscillations as $x\rightarrow +\infty$ exponentially decreasing amplitude but accelerating frequency,
\begin{align}
\rfun{\BesselF_{\mi k}}{\me^{x}}\simeq \sqrt{\frac{2}{\pp}}\me^{-x/2}\left\{\rfun{\cos}{\me^x-\frac{\pp}{4}}+O(\me^{-x})\right\},\label{asymptoticF}\\
\rfun{\BesselG_{\mi k}}{\me^{x}}\simeq \sqrt{\frac{2}{\pp}}\me^{-x/2}\left\{\rfun{\sin}{\me^x-\frac{\pp}{4}}+O(\me^{-x})\right\}\label{asymptoticG}
\end{align}
Thanks to this behavior both functions should naively contribute to the continuous spectrum. Using the method from \cite{szmytkowski2010} one can obtain the following orthogonality relations,
\begin{align}
\int_{-\infty}^{+\infty} \dif x\,
\rfun{\BesselF_{\mi k}}{\me^x} \rfun{{\BesselF}_{\mi l}}{\me^x} &=
\frac{1}{k}\rfun{\tanh}{\frac{\pp k}{2}}
\rfun{\delta}{k-l}, \\
\int_{-\infty}^{+\infty} \dif x\,
\rfun{\BesselG_{\mi k}}{\me^x} \rfun{\BesselG_{\mi l}}{\me^x} &=
\frac{1}{k}\rfun{\coth}{\frac{\pp k}{2}}
\rfun{\delta}{k-l}.
\end{align}
However both of these functions $\rfun{\BesselF_{\mi k}}{\me^x}$ and $\rfun{\BesselG_{\mi l}}{\me^x}$ can not be included into the continuous spectrum of a self-adjoint operator simultaneously as they are not orthogonal even when $k\neq l$ is different. Nevertheless, we note that their symmetrized scalar product vanishes,
\begin{equation}
\int_{-\infty}^{+\infty} \dif x\,\Big[
\rfun{\BesselF_{\mi k}}{\me^x} \rfun{\BesselG_{\mi l}}{\me^x}+\rfun{\BesselG_{\mi k}}{\me^x} \rfun{\BesselF_{\mi l}}{\me^x}\Big] =0.
\end{equation}
For negative energies $E\leq 0$ one naively obtains the continuous spectrum of square-integrable solutions,
\begin{equation}
\tilde{\psi}_\mu = \sqrt{2\mu}\rfun{\BesselJ_{\mu}}{\me^x},\quad E=-\mu^2.
\end{equation}
Similarly to the part of the spectrum with $E>0$, not all of these wavefunctions can be included into the spectrum of a self-adjoint operator because they are not orthogonal for different values of $\mu$ in general \cite{watson1922},
\begin{equation}
\int_{-\infty}^{+\infty}\dif x\, \rfun{\BesselJ_{\mu}}{\me^x}\rfun{\BesselJ_{\nu}}{\me^x}=2\frac{\sfun{\sin}{\frac{\pp}{2}(\mu-\nu)}}{\pp (\mu^2-\nu^2)}
\end{equation}

These peculiarities are caused by the operator $\hat{H}$, as defined on the standard domain of $\hat{p}^2$, being not essentially self-adjoint. Thus it actually describes a family of different self-adjoint extensions that are indistinguishable on sufficiently localized smooth functions but generate different unitary evolutions. Since this important topic is often neglected in the quantum mechanics courses we elucidate few important facts here and refer to \cite{Gitman2012,hall2013,bonneau2001} for details.

In infinite dimensional Hilbert spaces it is too restrictive to demand that the domain of the operator $\mathcal{D}(\hat{A})$ covered the whole Hilber space $\mathcal{H}$. Therefore operators including observables are usually defined on the domains that are merely \emph{dense} in $\mathcal{H}$ i.e. any element in the Hilbert space can be obtained as a limit of some sequence of elements in $\mathcal{D}(\hat{A})$. For example the operator $\hat{p}^2$ can not be defined on the whole $L^2(\mathbb{R})$ but is symmetric on the domain of all `bumps' - infinitely differentiable functions with compact support, $\mathcal{C}_c^{\infty}$.

However this leads to the following pitfall. Even if its domain is dense a \emph{symmetric operator} $\hat{A}$ such that,
\begin{equation}
(\psi,\hat{A}\chi)=(\hat{A}\psi,\chi),\quad\forall \psi,\chi\in \mathcal{D}(\hat{A}),
\end{equation}
does not in general possess important properties like spectral theorem and reality of eigenvalues. For $\hat{A}$ to be \emph{self-adjoint} its \emph{adjoint} $\hat{A}^\dagger$ defined as,
\begin{equation}
(\psi,\hat{A}^\dagger\chi)=(\hat{A}\psi,\chi),
\end{equation}
should have the same domain $\mathcal{D}(\hat{A}^\dagger)=\mathcal{D}(\hat{A})$. However in general the domain of $\hat{A}^\dagger$ is larger than the domain of $\hat{A}$. In many cases this happens because $\mathcal{D}(\hat{A})$ is selected to be too small and it is possible to find the self-adjoint operator called \emph{self-adjoint extension} on a larger domain that equals to $\hat{A}$ on the original domain. If such extension is unique $\hat{A}$ is said to be \emph{essentially self-adjoint}. But in general the operator $\hat{A}$ has many self-adjoint extenstions. This should not be considered as a pathology, rather the original definition of $\hat{A}$ happens to be incomplete and provides merely a local description of many different self-adjoint operators each generating its own unitary evolution.

For non-singular potentials bounded from below the Hamiltonian is essentially self-adjoint. However this is not a case for Eq.~\eqref{HamilApp}. It shows itself in the existence of square-integrable solutions of Eq.~\eqref{eq:negliouvschroed} with complex $E$. For example for $E_{\pm}=\pm 2\mi$ one gets,
\begin{equation}
\psi_{\pm}=C_{\pm} \rfun{\BesselJ_{1\pm \mi}}{\me^x}+\tilde{C}_{\pm} \rfun{\BesselJ_{-1\mp \mi}}{\me^x},
\end{equation} 
The dimensions of the subspaces of solutions corresponding to complex $E$ with $\operatorname{Im} E>0$ and $\operatorname{Im} E<0$ are known as deficiency indices $n_{+}$  and $n_{-}$ respectively. If $n_{+}=n_{-}=0$ (i.e. there are no such solutions) the operator is essentially self-adjoint, that is its self-adjoint extension is unique. If $n_{+}\neq n_{-}$ no self-adjoint extension exists. In our case the square-integrability requires $\tilde{C}_{\pm}=0$ however $C_{\pm}\neq 0$ is allowed. Therefore $n_{+}=n_{-}=1$. According to the Weyl--von Neumann theorem \cite{bonneau2001} this means that a single parameter family of self-adjoint extensions exists.

The functions $\tilde{\psi}_\mu$ are square integrable but don't belong to $\mathcal{C}_c^{\infty}$. As result the $\hat{p}^2$ and $\hat{H}$ are not generally symmetric on these solutions,
\begin{equation}\begin{split}
&\int_{-\infty}^{+\infty}\dif x\, \tilde{\psi}_\mu^\ast(x)\Big[\hat{H}\tilde{\psi}_\nu(x)\Big] 
	- \int_{-\infty}^{+\infty}\dif x\, \Big[\hat{H}\tilde{\psi}_\mu(x)\Big]^\ast\tilde{\psi}_\nu(x) \\
= & \frac{2}{\pp}\sqrt{\mu\nu}\sfun{\sin}{\frac{\pp}{2}(\mu-\nu)}.
\end{split}
\end{equation}
To extend the domain of $\hat{H}$ conserving its symmetricity we consider the new functional class bigger than $\mathcal{C}_c^{\infty}$ with a specific oscillatory behavior as $x\rightarrow+\infty$,
\begin{equation}
\psi\sim \me^{-x/2}\rfun{\cos}{\me^x-\frac{\pp}{2}a-\frac{\pp}{4}},
\end{equation}
where $a$ is an arbitrary parameter $a\in[0,2)$. For $E>0$ using Eqs.~\eqref{asymptoticF},\eqref{asymptoticG} we then get non-degenerate continuous spectrum,
\begin{equation}
\rfun{\Xi_{k}^{(a)}}{x}=\mathcal{N}_k^{(a)}\left[\rfun{\BesselF_{\mi k}}{\me^x}\cos\frac{\pp a}{2}+\rfun{\BesselG_{\mi k}}{\me^x}\sin\frac{\pp a}{2}\right].\label{eq:negliouvcont}
\end{equation}
\begin{equation}\begin{split}
\left(\mathcal{N}_k^{(a)}\right)^{-2}=\frac{1}{k^2}\rfun{\tanh^2}{\frac{\pp k}{2}}&\cos^2\frac{\pp a}{2}+\\
+&\frac{1}{k^2}\rfun{\coth^2}{\frac{\pp k}{2}}\sin^2\frac{\pp a}{2},
\end{split}
\end{equation}
whereas for $E\leq 0$ using 10.7.2 from \cite{olver2010} we obtain the discrete spectrum,
\begin{equation}
\Phi^{(a)}_{n}(x)=\sqrt{2(2n+a)}\rfun{\BesselJ_{2n+a}}{\me^x},\quad E=-(2n+a)^2.\label{eq:negliouvdisc}
\end{equation}
The resulting full spectrum forms orthonormal set,
\begin{equation}\begin{split}
&\int_{-\infty}^{+\infty}\dif x\,\Big[\Phi^{(a)}_{n}(x)\Big]^\ast\Phi^{(a)}_{m}(x)=\delta_{nm},\\
&\int_{-\infty}^{+\infty}\dif x\,\Big[\Phi^{(a)}_{n}(x)\Big]^\ast\Xi^{(a)}_{k}(x)=0,
\end{split}
\end{equation}
\begin{equation}
\int_{-\infty}^{+\infty}\dif x\,\Big[\Xi^{(a)}_{k}(x)\Big]^\ast\Xi^{(a)}_{l}(x)=\delta(k-l).
\end{equation}

It is interesting that the discreteness of the spectrum for $E<0$ and the non-degeneracy of the continuous spectrum for $E>0$ makes the abyss of the potential at large positive $x$ analogous to a reflecting wall. The classical trajectories for the particle described by $H$ reach infinity in finite time. Therefore in the first WKB approximation, the Gaussian wave packet also reaches the infinity in finite time. The subsequent motion of the particle may be described as a bounce from infinity. The non-uniqueness of the self-adjoint extension for $\hat{H}$ may be understood intuitively in the following way. After crossing over infinity the wave function may be multiplied by an arbitrary phase factor $\me^{2\pp\mi a}$ without losing the conservation of probability. Thus we have a family of unitary evolution operators generated by different self-adjoint extensions of $\hat{H}$ that locally are indistinguishable however differ at finite times.

Another way, perhaps more physical, to understand this non-uniqueness is to consider the regularized potential, for example introducing an infinitely high wall at $x=L$ that forms a potential well with the fall of the potential at large $x$ playing the role of another wall,
\begin{equation}
\Big[-\partial_x^2-\me^{2x}\Big]\psi=E\psi,\quad \psi\Big\vert_{x=L}=0.
\end{equation}
Even in the limit $L\rightarrow+\infty$ the energy levels for $E<0$ stay apart from each other and the spectrum remains to be discrete. The non-uniqueness of the self-adjoint extension takes the form of the regularization-dependence. The parameter $a$ can be shown to be equal to,
\begin{equation}
\frac{\pp a}{2}=\left(\me^L-\frac{3\pp}{4}\right)\mod \pp
\end{equation}

\section{Classical solutions of Liouville cosmology}
\label{sec:classical}

Consider a Friedmann--Lema\^{i}tre model minimally coupled with a
spatially isotropic and homogeneous
Liouville field. The Friedmann--Lema\^{i}tre--
Robertson--Walker
(FLRW)
metric is
\begin{equation}
\dif s^2 = \rfun{N^2}{t}\,\dif t^2 -\me^{2\alpha(t)}\,\dif {\vec x}^2,
\end{equation}
where $N(t)$ is the lapse function, and $a(t)= \exp \alpha(t)$ the cosmological
scale factor; moreover, the scalar field is a function only of time, $\phi =
\phi(t)$. With $\kap = 8 \pp G$, $\sigma = \pm 1$ and $\lambda \in
\mathbb{R}$, the minisuperspace action reads
\begin{equation}\label{eq:action}
S(\lambda)=\int \dif t\,
    N \me^{3\a}\left(
        -\frac{3}{\kap}\frac{\dot\a^2}{N^2}
        +\sigma\frac{\dot\phi^2}{2N^2}
        -V\me^{\lambda\phi}
    \right),
\end{equation}
where $\sigma = +1$ gives a \emph{quintessence model} \cite{caldwell1998},
and $\sigma = -1$ is dubbed as a \emph{phantom model} \cite{caldwell2002}. From Eq.~\eqref{eq:action}
one readily derives the Hamiltonian density
\begin{equation}\label{eq:Hamiltonian}
\rfun{\mathcal{H}}{\lambda} =  N \me^{-3\a}\left(
    -\frac{\kap}{12}p_\a^2
    +\sigma\frac{1}{2} p_\phi^2
    +V \me^{6\a+\lambda\phi}
\right),
\end{equation}
in terms of $\a$ and $\phi$, as well as their canonical momenta
\begin{equation}
p_\a =
        -\me^{3\a}\frac{6}{\kap}\frac{\dot\a}{N},
\quad
p_\phi = \sigma \me^{3\a}\frac{\dot\phi}{N};
\end{equation}
the significance of stressing $\lambda$ will be elaborated in Sec.~\eqref{sec:limit}.
It has been shown in \cite{lanchen2016,andrianov2015} that
 \begin{equation}\label{eq:IntegralofMotion}
\omega \coloneqq \me^{3\a}
\left(
    \frac{\lambda}{\kap}\frac{\dot\a}{N}
    +\sigma\frac{\dot\phi}{N}
\right)
\end{equation}
is an integral of motion, i.e.\ $\omega$ is a constant on the constraint surface
\begin{equation}
\dot \omega =-\frac{2}{\lambda } \mathcal{H} \approx 0,
\end{equation}
where $\approx$ represents Dirac's weak equivalence
\cite{dirac1964,gitman1990,rothe2010,henneaux1994,prokhorov2009}. 
Applying Eq.~\eqref{eq:IntegralofMotion}
to the Friedmann equation
\begin{equation}
\frac{\dot\a^2}{N^2}
=\frac{\kap}{3}\left(
       \sigma \frac{\dot\phi^2}{2N^2}
        +V\me^{\lambda\phi}
\right),
\end{equation}
one can eliminate the lapse function $N$ and obtain a non-linear equation
\begin{equation}\label{eq:this}
\left( \frac{\dif\a(\phi)}{\dif \phi}\right)^2-\sigma\frac{\kap}{6}=
    \frac{3\omega^2}{\kap V}\me^{6\a(\phi)+\lambda\phi}
    \left(
        \frac{\lambda}{\kap}\frac{\dif\a(\phi)}{\dif \phi}
        +\sigma
    \right)^2
\end{equation}
in terms of minisuperspace variables $\alpha$ and $\phi$ only, and
$\dot\a/\dot\phi$ has already been replaced by
$\dif\a(\phi)
/ \dif\phi$. Eq.~\eqref{eq:this} can be solved with the help of a change of variables
\begin{equation}
x \text{ (or $y$) } \coloneqq 6\a+\lambda\phi.
\label{eq:x=6alpha+}
\end{equation}
where $x$ is for quintessence and $y$ is for phantom.

Defining
\begin{equation}
m_x \coloneqq -6\varkappa+\lambda^2,
\label{eq:m_x-def}
\end{equation}
the solution for a quintessence model $\sigma=+1$ can be divided into two cases:
\begin{enumerate}
\item When $m_x$ and $V$ are of different sign, one obtains
\begin{equation}\label{eq:classicalquintessence}
\me^{6 \a +\lambda  \phi }
	=\frac{3 \kap \omega^2}{- V m_x}
		\rfun{\csch^2}{\lambda\sqrt{\frac{3}{2 \kap }} \a
			+\sqrt{\frac{3 \kap}{2}} \phi
				+c_1},
\end{equation}
where $c_1$ is an integration constant associated with the initial conditions.
Eq.~\eqref{eq:classicalquintessence} contains two distinct solutions separated by
$\lambda \sqrt{\frac{3}{2 \kap }} \a +\sqrt{\frac{3 \kap}{2}} \phi +c_1 = 0$
due
to the divergence of $\csch x$ for $x \to 0$. Both of the solutions can be
interpreted as an expansion model, see e.g.\ Fig.~\eqref{fig:sinh-F}.
For $\omega=0$, one recovers the power-law special solution or $\a\propto \phi$
in \cite{dabrowski2006}.

\item When $m_x$ and $V$ are of the same sign, one has
\begin{equation}
\me^{6 \a +\lambda  \phi }
	=\frac{3\kap  \omega^2}{ V m_x}
		\rfun{\sech^2}{\lambda\sqrt{\frac{3}{2 \kap }} \a
			+\sqrt{\frac{3 \kap}{2}} \phi
				+c_1},
\label{eq:classicalquintessence2}
\end{equation}
this trajectory contains a single turning point in finite domain of
minisuperspace.
\end{enumerate}
As for the second case the quantization is straightforward we will concentrate on the first case.

\begin{figure}
\begin{center}
\includegraphics[width=0.8\linewidth]{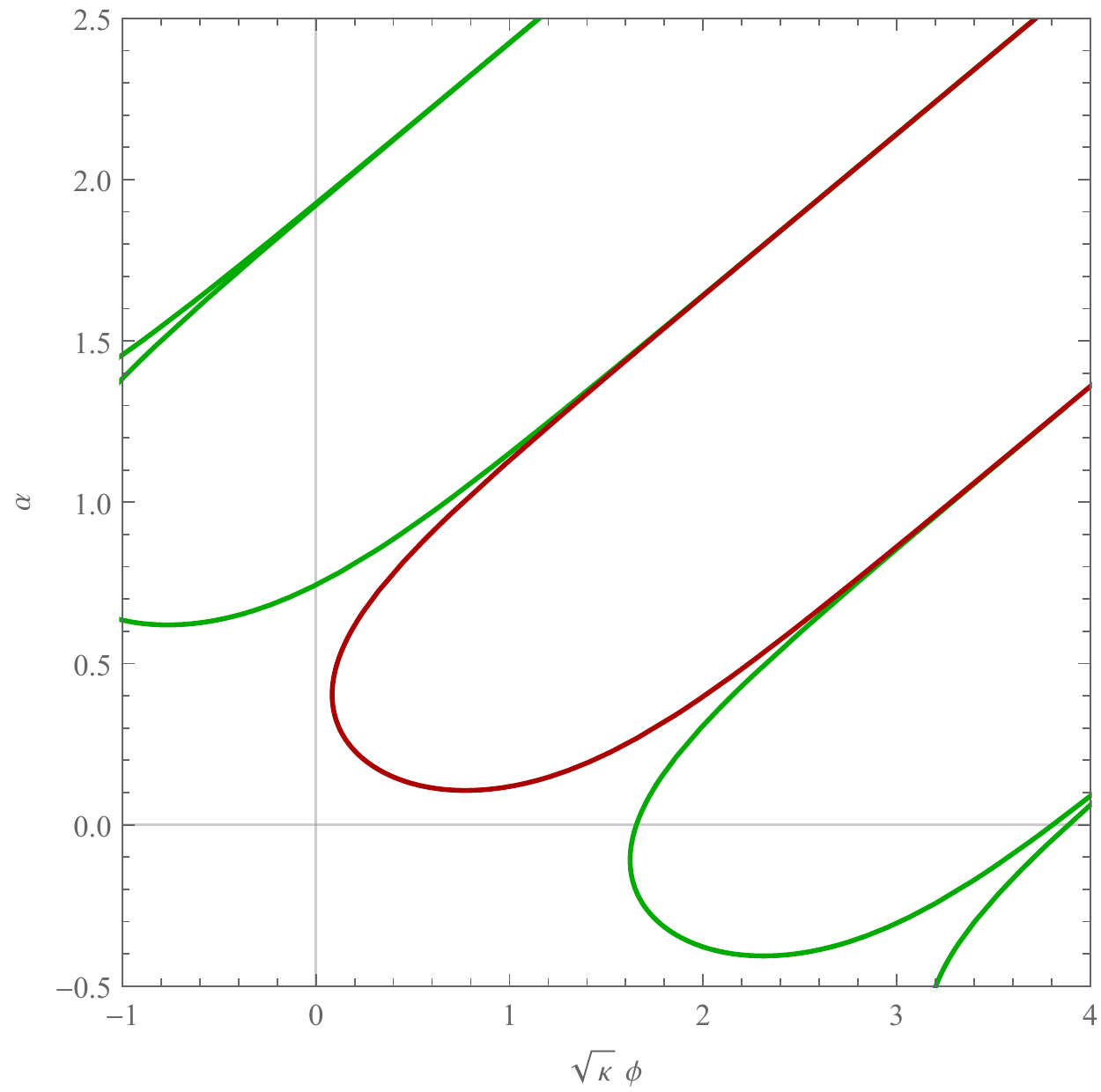}
\end{center}
\caption{Multiple solutions contained in Eq.~\eqref{eq:classicalphantom} for
$\varkappa = 1$, $V = 1$, $\lambda = 2$ and $A^2 = 160/3$. Only one of them is
physical, which can be taken to be the red one; the other trajectories (in
green) appear due to the ambiguity in the timeless Eq.~\eqref{eq:classicalphantom}.
If a time parameter is chosen, the other branches will disappear.
\label{fig:sec}}
\end{figure}

Similar to Eq.~\eqref{eq:m_x-def}, one can define
\begin{equation}
m_y \coloneqq +6 \varkappa + \lambda^2 > 0
\label{eq:m_y-def}
\end{equation}
for phantom model with $\sigma=-1$. The solution reads
\begin{equation}\label{eq:classicalphantom}
\me^{6\a+\lambda\phi}=
\frac{3\kap \omega^2}{ V m_y}\rfun{\sec^2}{
	\lambda\sqrt{\frac{3}{2\kap}}\a
  	  - \sqrt{\frac{3\kap}{2}}\phi+c_2},
\end{equation}
where $c_2$ is another integration constant. Eq.~\eqref{eq:classicalphantom}
contains
a infinite family of distinct solutions separated by two types of cosmological
singularities at infinity, due to the periodic divergences of $\sec$ function,
see Fig.~\eqref{fig:sec}.

With a given set of initial conditions, the universe runs only along one branch.
Hence the domain of minisuperspace variables in Eq.~\eqref{eq:classicalphantom} has
to be restricted
\begin{equation}\label{eq:additional}
-\frac{\pp}{2}+k\pp
	\le \lambda\sqrt{\frac{3}{2\kap}}\a
  	  - \sqrt{\frac{3\kap}{2}}\phi+c_2
	\le \frac{\pp}{2}+k\pp,\quad k\in \mathbb{Z},
\end{equation}
such that only one trajectory between a pair of singularities is selected.
In other words, eliminating time parameter in the classical solution gives rise
to Eq.~\eqref{eq:classicalphantom} that covers redundant trajectories as well,
which should be eliminated by the additional condition Eq.~\eqref{eq:additional}.
For simplicity, one can choose $c_1 \equiv 0$, $k \equiv 0$ and obtain
\begin{equation}\label{eq:periodicity-boundary}
-\frac{\pp}{2} \le \lambda\sqrt{\frac{3}{2\kap}}\a
  	  - \sqrt{\frac{3\kap}{2}}\phi
\le \frac{\pp}{2},
\end{equation}
which could be applied as a boundary condition in quantum theory. According to
DeWitt's criterion \cite{dewitt1967} , the wave functions must
vanish at classical singularities. This restriction on the classical
domain of variables affords the possibility to determine the ambiguity of self-adjoint extension,
if one prefers to fix the periodicity of wave function with respect to $\tau$.
see Sec.~\eqref{sec:quantization}.

\section{Dirac quantization of Liouville cosmology}
\label{sec:quantization}

\subsection{Inner product and probabilities}

On of the basic building blocks of any quantum model is the inner product that allows to assign probabilities. However this is a long standing problem in quantum
cosmology due to the Wheeler--DeWitt equation being of the Klein--Gordon type. The naturally conserved Klein--Gordon inner product corresponds to the indefinite norm
\cite[ch.~5]{kiefer2012}. Pseudo-Hermitian quantum mechanics
\cite{mostafazadeh2010} provides a cure and will be applied here to
reconstruct wave packets based on consistent norms.

Assume that the Wheeler--DeWitt equaton can be written in the form,
\begin{equation}
\partial_\tau^2 \psi + \mathbf{D} \psi = 0,
\label{eq:KG-Mos}
\end{equation}
The simplest approach is to use the usual
Schr\"odinger inner product,
\begin{equation}\label{eq:naive-L2-norm}
\rbr{\psi_1, \psi_2}_\text{S} \coloneqq \int_{-\infty}^{+\infty}\dif x\,
\rfun{\psi_1^*}{\tau, x}\rfun{\psi_2}{\tau, x},
\end{equation}
however it is not conserved. On the other hand, the naturally conserved Klein--Gordon inner product,
\begin{equation}
\begin{aligned}\rbr{\psi_1, \psi_2}_\text{KG} \coloneqq \int_{-\infty}^{+\infty}\dif x\,
\Big(&\rfun{\dot{\psi_1^*}}{\tau, x}\rfun{\psi_2}{\tau, x}\\
&-\rfun{\psi_1^*}{\tau, x}\rfun{\dot{\psi_2}}{\tau, x}\Big),\label{eq:KG-norm}\end{aligned}
\end{equation}
is not suitable to define the probabilities as it is not positive-definite.

In the pseudo-Hermitian quantum mechanics, an alternative definition of inner
product by Mostafazadeh can be adapted from
\cite{mostafazadeh2002,mostafazadeh2006}, where a family of Hilbert spaces
with a corresponding pseudo-Hamiltonian were constructed for the Klein--Gordon
equation~\eqref{eq:KG-Mos}.
The solution of Eq.~\eqref{eq:KG-Mos} are
endowed with, again, the Schr\"odinger $\rfun{L^2}{\mathbb{R}}$ inner product in
Eq.~\eqref{eq:naive-L2-norm}, and $\mathbf{D}$ (not necessarily independent of
$\tau$!) is required to be Hermitian with eigenfunctions and
non-negative eigenvalues
\begin{equation}
\mathbf{D}\psi_n = \nu_n^2 \psi_n.
\end{equation}
The Mostafazadeh inner product of the new Hilbert space, which features
time-translational invariance with respect to $\tau$, can be chosen to be
\begin{equation}
\rbr{\psi_1, \psi_2}_\text{M} \coloneqq \frac{1}{2\mu}
\sbr{\rbr{\psi_1, \mathbf{D}^{+1/2} \psi_2}_\text{S} +
\rbr{\dot\psi_1, \mathbf{D}^{-1/2} \dot\psi_2}_\text{S}},
\label{eq:inner-prod-Mosta}
\end{equation}
in which $\mu$ is a normalizing constant, $\dot\psi\coloneqq\partial_\tau\psi$,
and $\mathbf{D}^\gamma$ is defined by the spectral decomposition
\begin{equation}
\mathbf{D}^\gamma \coloneqq \sum_n \nu_n^{2\gamma} \mathbf{P}_n,\quad
\mathbf{P}_n \psi \coloneqq \psi_n \rbr{\psi_n, \psi}_\text{S}.
\end{equation}

Eq.~\eqref{eq:inner-prod-Mosta} is manifestly positive-definite, but its
integrand $\varrho$ is, in general, complex. Luckily, a non-negative density
\begin{equation}
\rho \coloneqq \frac{1}{2\mu} \sbr{\vbr{\mathbf{D}^{+1/4}\psi}^2 +
\vbr{\mathbf{D}^{-1/4}\dot\psi}^2}
\end{equation}
can be defined whose integral gives the Mostafazadeh inner product
\begin{equation}
\int \dif \vec{x}\,\rfun{\rho}{\vec{x}} \equiv \rbr{\psi, \psi}_\text{M}
\eqqcolon \int \dif \vec{x}\,\rfun{\varrho}{\vec{x}}.
\end{equation}
Therefore $\rho$ is a good candidate for a probability density in the
minisuperspace.

\subsection{Quintessence field}
In addition to Eq.~\eqref{eq:x=6alpha+}, a further transformation
\begin{equation}
\tau \coloneqq \frac{6\kap}{m_x}\left(\frac{\lambda}{\kap}\a+\phi\right),
\end{equation}
is to be performed in order to separate the variables, which is related to
$\omega$ by
\begin{equation}\label{eq:def-omega}
\omega =\frac{1}{\tilde N} \frac{\dif \tau}{\dif t},\qquad
\tilde N \coloneqq \frac{6\kap}{m_x} N \me^{-3\a}.
\end{equation}
Because of Eq.~\eqref{eq:def-omega}, $\tau$ can be treated as the \emph{time of a
Klein--Gordon-type equation} and $\omega$ as its Fourier conjugate. The
Hamiltonian in Eq.~\eqref{eq:Hamiltonian} then becomes
\begin{equation}
\mathcal{H}_x = N \me^{\frac{3 \kap x}{m_x}-\frac{\lambda \tau }{2}}
\left(
-\frac{3\kap}{m_x} p_\tau^2
+\frac{m_x}{2}p_x^2
+V \me^x
\right),
\end{equation}
which is of Klein--Gordon form. Promoting the canonical variables to operators
in the position representation
\begin{equation}
\tau \to \tau, \quad x \to x;\qquad
p_\tau \to -\mi \hslash \frac{\partial}{\partial \tau},\quad
p_x \to -\mi \hslash  \frac{\partial}{\partial x},
\end{equation}
one can obtain the Wheeler--DeWitt equation
\begin{equation}
\left(
\frac{3\kap\hslash^2}{m_x} \partial_\tau^2
-\frac{m_x\hslash^2}{2}\partial_x^2
+V \me^x
\right)
\rfun{\varPsi}{\tau,x}=0,
\end{equation}
which is Eq.~\eqref{eq:KG-Mos} with
\begin{equation}
\mathbf{D}=- \frac{\hslash^2 m_x^2}{6\varkappa} \partial_x^2 +
\frac{V m_x}{3\varkappa} \me^x
\label{eq:def-DD1}
\end{equation}
Its solution can be represented by the Fourier integral
\begin{equation}\label{eq:integral}
\rfun{\varPsi}{\tau,x}=\int _{-\infty}^{+\infty} \dif \omega\,
\rfun{\mathcal{A}}{\omega}
	\me^{-\frac{\mi}{\hslash}\tau \omega}
	\rfun{\psi}{\omega,x},
\end{equation}
where $\rfun{\psi}{\omega,x}$ satisfies
\begin{equation}\label{eq:quintessence}
\left(
-\frac{ 3\kap \omega^2}{m_x}
-\frac{m_x\hslash^2}{2}\partial_x^2
+V \me^x
\right)\rfun{\psi}{\omega,x}=0.
\end{equation}

In order to save the Hermiticity of operators and define meaningful probability densities \cite{barvinsky2014, kiefer2012} one can demand the solution to be square integrable. The expectation value of a physical observable can be defined naively by
\begin{equation}
\braket{\mathcal{O}} = \rbr{\varPsi, \hat{\mathscr{O}} \varPsi}_\text{S}
\coloneqq \int \dif x\, \rfun{\varPsi^*}{\tau,x} \hat{\mathscr{O}}
\rfun{\varPsi}{\tau,x}.
\end{equation}

When $m_x V>0$ the operator is essentially self-adjoint and the quantization proceeds in a straightforward fashion. In contrast when $m_x V<0$, the Eq.~\eqref{eq:quintessence} can be regarded as
the stationary Schr\"odinger equation with negative potential unbounded from below
  and the corresponding operator is not essentially self-adjoint which is the problem that was considered in detail in section \ref{sec:unstable}.
The square-integrable functions can be represented as superpositions of eigenfunctions of ,
\begin{align}
\rfun{\varPsi}{\tau,x}=&\int _{-\infty}^{+\infty} \dif \omega\,
\rfun{\mathcal{A}}{\omega}
	\me^{-\frac{\mi}{\hslash}\tau \omega}
	\rfun{\Xi_{\nu}^{(a)}}{\frac{2}{\hslash}\sqrt{\frac{-2V}{m_x}}\me^{x/2}}\nonumber\\
&+\sum_{(\pm)}\sum_{n=0}^{+\infty} \mathcal{C}_n^{(\pm)}\me^{\pm\frac{1}{\hslash}\tau |\omega_n|} \rfun{\Phi_n^{(a)}}{\frac{2}{\hslash}\sqrt{\frac{-2V}{m_x}}\me^{x/2}},\label{eq:wavequintessence2}
\end{align}
where $\nu$ is given by,
\begin{equation}
\nu=2\frac{\sqrt{6\kap}}{\hslash }\abs{\frac{\omega}{m_x}}.
\label{eq:index}
\end{equation}
and the functions $\Xi_\nu^{(a)}$ and $\Phi_n^{(a)}$ are defined in Eq.~\eqref{eq:negliouvcont} and Eq.~\eqref{eq:negliouvdisc} respectively. 
The solution contains arbitrary parameter $a\in[0,2)$ 
specifying the self-adjoint extension. 
The first part of the wave function corresponds to the solution of 
Eq.~\eqref{eq:quintessence} with positive $\omega^2>0$,
while the second is derived from same equation with negative $\omega^2<0$.
The discrete purely imaginary $\omega_n$
\begin{equation}
2n+a=2\frac{\sqrt{6\kap}}{\hslash }\abs{\frac{\omega_n}{m_x}},
\end{equation}
are required for completeness and hermiticity,
but they produce growing and decreasing modes,
which are not compatible with conservation neither of the Klein--Gordon norm Eq.~\eqref{eq:KG-norm} nor of the Mostafazadeh norm Eq.~\eqref{eq:inner-prod-Mosta}. It is worth noting that these modes also violate the classical restriction $\omega^2>0$
imposed by reality of metric and field variables in Eq.~\eqref{eq:classicalquintessence2}.
It will be shown below that the wave packets along the correct classical trajectories
can be constructed only from the continuous spectrum $\Xi_{\mi\nu}^{(a)}$.
Thus we conclude that on the physical space there's no contribution from the discrete spectrum, i.e.
\begin{equation}
\mathcal{C}_n^{(\pm)}=0
\end{equation}
as a result in the quantum model both unitary evolution and correct classical limit can be guaranteed.

\subsection{Phantom field}
A transformation similar to the quintessence case
\begin{equation}
\tau \coloneqq \frac{6\kap}{m_y}\left(\frac{\lambda}{\kap}\a-\phi\right)
\end{equation}
can be made for phantom, such that the Hamiltonian in Eq.~\eqref{eq:Hamiltonian}
becomes
\begin{equation}
\mathcal{H}_y=N \me^{-\frac{\lambda \tau}{2} - 3 \frac{\kap y}{m_y}}
\rbr{-\frac{3 \kap }{ m_y}p_{\tau }^2 -\frac{m_y}{2}  p_y^2+V \me^y}.
\end{equation}
The Wheeler--DeWitt equation then reads
\begin{equation}
\left(
\frac{3\kap\hslash^2}{m_y} \partial_\tau^2
+\frac{m_y\hslash^2}{2}\partial_y^2
+V \me^y
\right)
\varPsi(\tau,y)=0,
\end{equation}
That again takes the form of Eq.~\eqref{eq:KG-Mos} with,
\begin{equation}
\mathbf{D}=
+ \frac{\hslash^2 m_y^2}{6\varkappa} \partial_y^2 +
\frac{V m_y}{3\varkappa} \me^y.
\label{eq:def-DD2}
\end{equation}

The separation of variables allows us to find the solution using two equations
\begin{equation}\label{eq:phantom-2}
\rbr{\partial_\tau^2+\omega^2}\rfun{f}{\tau,\omega}=0.
\end{equation}
and
\begin{equation}\label{eq:phantom-1}
\left(
\frac{m_y\hslash^2}{2}\partial_y^2
+V \me^y
\right)
\rfun{\psi}{\omega,y}=\frac{3\kap\omega^2}{m_y}\rfun{\psi}{\omega,y},
\end{equation}
The equation Eq.~\eqref{eq:phantom-1} is very similar to Eq.~\eqref{eq:quintessence} with $m_x V<0$,
hence it will give rise to a similar problem, which will be considered in \eqref{sec:unstable} as well.
But in this case the sign of $\omega^2$ is different with quintessence, 
the general solutions include two parts, one is the time-oscillating functions constructed from the discrete spectrum, and the other is decreasing and increasing functions as the superpositions of the modes with continuous spectrum,
\begin{align}
\rfun{\varPsi}{\tau,y}=&\sum_{(\pm)}\sum_{n=0}^{+\infty} \mathcal{A}_n^{(\pm)}\me^{\mp\frac{i}{\hslash}\tau \omega_n}
 \rfun{\Phi_n^{(a)}}{\frac{2}{\hslash}\sqrt{\frac{2V}{m_y}}\me^{y/2}}\nonumber\\
&+\int _{-\infty}^{+\infty} \dif \tilde{\omega}\,
\rfun{\mathcal{B}}{\tilde{\omega}}
	\me^{-\frac{1}{\hslash}\tau \tilde{\omega}}
	\rfun{\Xi_{\nu}^{(a)}}{\frac{2}{\hslash}\sqrt{\frac{2V}{m_y}}\me^{y/2}},\label{eq:phantom-sol-full}
\end{align}
where,
\begin{equation}
2n+a=\frac{2 \sqrt{6 \kap}  }{\hslash  m_y}\omega_n,\quad	\mu = \frac{2 \sqrt{6 \kap}  }{\hslash  m_y}\tilde{\omega}.
\end{equation}
Similarly to the case of quintessence with $m_x<0$ and $V>0$ one can exclude the continuous spectrum to preserve both probability conservation and correct classical limit with $\omega^2>0$ by setting
\begin{equation}
\rfun{\mathcal{B}}{\tilde{\omega}}=0.
\end{equation}
The resulting wave packet can be written explicitly as,
\begin{equation}\label{eq:phantom-sol}
\rfun{\varPsi}{\tau,y}=
\sum_{(\pm)}\sum_{n=0}^{+\infty} \mathcal{A}_n^{(\pm)}\sqrt{2n+a} \me^{\mp\frac{\mi}{\hslash}\tau \omega_n} \rfun{\BesselJ_{2n+a}}{\frac{2}{\hslash}\sqrt{\frac{2V
}{m_y}}\me^{y/2}}
\end{equation}
If the wave packet is only restricted to the positive frequencies the discreteness will be associated with periodicity of $\tau$. The value of $a$ can be fixed by the condition,
\begin{equation}\label{eq:periodicity}
\rfun{\varPsi}{\tau,y}=\me^{\mi a \pp} \rfun{\varPsi}{\tau+\frac{ 2\pp\sqrt{6 \kap}  }{m_y },y}
\end{equation}
Such periodic condition also guarantees the self-adjointness of the $\partial_\tau^2$ operator. If both positive and negative frequencies are included the only possibilities are $a=0$ and $a=1$ corresponding to periodic and antiperiodic wavefunctions respectively.

\section{The limit $\lambda\to 0$}
\label{sec:limit}

As a verification of our approach to the minisuperspace trajectory, the limit
$\lambda \to 0$ will be considered, which have been extensively studied as a
pedagogic model, see e.g.\ \cite{barvinsky2014,dabrowski2006,kamenshchik2017b}.
This limit enforces $m_x<0$ which will be assumed for the rest of the section.

The action in Eq.~\eqref{eq:action} in this limit becomes
\begin{equation}
\rfun{S}{0}=\int \dif t\,
    N \me^{3\a}\left(
        -\frac{3}{\kap}\frac{\dot\a^2}{N^2}
        +\sigma\frac{\dot\phi^2}{2N^2}
        -V
    \right),
\end{equation}
and the integral of motion Eq.~\eqref{eq:IntegralofMotion} tends to
\begin{equation}
\omega\rightarrow \sigma\me^{3\a}
\frac{\dot\phi}{N}\equiv p_\phi, \qquad \lambda\to 0.
\end{equation}
For quintessence model with $V<0$, one obtains the classical
solution from Eq.~\eqref{eq:classicalquintessence} by setting $\lambda=0$
\begin{equation}
\me^{6 \a  }
	=\frac{ p_\phi^2}{2 V}
	\rfun{\csch^2}{\sqrt{\frac{3 \kap}{2}} \phi+c_1}.
\end{equation}
The quantum solution can also be calculated in similar way
\begin{equation}\label{eq:quintessencezerolambda}
\begin{split}
\varPsi&\left(\a,\phi\right)=\int_{-\infty}^{+\infty} \dif p_\phi\,
	\rfun{\mathcal{A}}{p_\phi}
	\me^{-\frac{\mi}{\hslash} \phi p_\phi}\times\\
		&\times\left[c_1\rfun{\BesselF_{\mi \nu}}{
		\frac{2}{\hslash}\sqrt{\frac{V}{3\kap}}\me^{3\a}}
		+c_2\rfun{\BesselG_{\mi \nu}}{
		\frac{2}{\hslash}\sqrt{\frac{V}{3\kap}}\me^{3\a}}
		\right]
\end{split}
\end{equation}
where the index of the Bessel function becomes
\begin{equation}
\nu=\sqrt{\frac{2}{3\kap}}\left|\frac{p_\phi}{\hslash}\right|.
\end{equation}
For the phantom model, one obtains
\begin{equation}
\me^{6\a}=\frac{ p_\phi^2}{ 2V}
	\rfun{\sec^2}{ \sqrt{\frac{3\kap}{2}}\phi+c_2},
\end{equation}
and
\begin{equation}\label{eq:phantom0}
\rfun{\varPsi}{\a,\phi}=\sum_n
\rfun{\mathcal{A}}{p_n}
\me^{-\frac{\mi}{\hslash} p_n \phi}
\rfun{\BesselJ_{2n+a}}{\frac{2}{\hslash}\sqrt{\frac{V}{3\kap}}\me^{3\a}}.
\end{equation}

\section{Semiclassical Wave Packets and Comparisons with Classical Solutions}
\label{sec:semiclassical}

With the explicit form of minisuperspace trajectories at hand, its comparison
with the quantum solutions becomes more transparent, since the latter does not
depend on any time parameter, but only the minisuperpace coordinates. It is
expected that a classical trajectory could be restored from the wave functions
at the limit $\hslash\to 0$, which must be consistent with the results in
Sec.~\eqref{sec:classical}; furthermore, the cosmological wave packets are expected
to go along the classical trajectories in minisuperspace, which can be
visualized in plots.

\subsection{WKB limit as $\hslash \to 0$}

The minisuperspace Wheeler--DeWitt wave functions can be compared with the
classical trajectories by taking the WKB limit, i.e.\ expanding at $\hslash\to
0$.

For the model with $m_x V<0$, it is sufficient to consider the phase contribution of
 $\rfun{\BesselF_{\mi \nu}}{x}$.
The uniform asymptotic expansion of unmodified
Bessel function for large index $\nu$ provides the leading order
\cite{dunster1990}
\begin{align}
\rfun{\BesselF_{\mi\nu}}{\nu z} &\sim \left(\frac{2}{\pp \nu}\right)^{1/2}
	 \rbr{1+z^2}^{-1/4}
	\rfun{\cos}{\zeta\nu -\frac{\pp}{4}}, \\
\zeta &\coloneqq \rbr{1+z^2}^{1/2}+\rfun{\ln}{\frac{z}{1+\rbr{1+z^2}^{1/2}}}.
\end{align}
The zeroth-order action reads
\begin{equation}
\frac{S_0}{\hslash} = -\frac{\tau\omega}{\hslash}
	+\nu \sbr{\rbr{1+z^2}^{1/2}+\rfun{\ln}{\frac{z}{1+\rbr{1+z^2}^{1/2}}}}
	-\frac{\pp}{4},
\end{equation}
where
\begin{equation}
\nu \coloneqq 2\frac{\sqrt{6\kap}}{\hslash }\abs{\frac{\omega}{m_x}},
\qquad
z \coloneqq \frac{1}{\abs{ \omega} }\sqrt{\frac{-V m_x}{3 \kap}}\me^{x/2}.
\end{equation}
$\partial S_0/\partial \omega=0$ gives
\begin{equation}
\me^{6 \a +\lambda  \phi }
	=\frac{3 \kap  \omega ^2}{ -m_x V}
		\rfun{\csch^2}{\lambda\sqrt{\frac{3}{2 \kap }} \a
			+\sqrt{\frac{3 \kap}{2}} \phi},
\end{equation}
which is consistent with Eq.~\eqref{eq:classicalquintessence2} up to a choice of
$c_2$.

For the phantom model, the Bessel function $\BesselJ_n$ is to be considered, whose
leading-order expansion reads
\begin{equation}
{\BesselJ}_{\nu}(\nu z)\sim
\rbr{\frac{4\zeta}{1-z^2}}^{1/4}
\frac{\rfun{\mathrm{Ai}}{\nu^{2/3}\zeta}}{\nu^{1/3}},
\end{equation}
for
\begin{align}
\frac{2}{3}(-\zeta)^{3/2} &=
    \rbr{z^2-1}^{1/2}-\arccos\frac{1}{z},
    \qquad \vbr{z}>1, \\
\rfun{\mathrm{Ai}}{\nu^{2/3}\zeta} &\sim
\frac{1}{\sqrt{\pp}(-\nu^{2/3}\zeta)^{1/4}} \rfun{\cos}{\frac{2}{3}\nu
(-\zeta)^{3/2}
-\frac{\pp}{4}}.
\end{align}
The zeroth-order of action then reads
\begin{equation}
\frac{S_0}{\hslash} = \frac{\tau\omega}{\hslash}+
	 \nu\sbr{\rbr{z^2-1}^{1/2}-\arccos\frac{1}{z}}-\frac{\pp}{4},
\end{equation}
where
\begin{equation}
\nu =\frac{2\sqrt{6\kap}}{\hslash m_y}\abs{\omega},\quad
 z=\frac{1}{\abs{\omega}}\sqrt{\frac{V m_y}{3\kap}} \me^{y/2}.
\end{equation}
Consequently $\partial S_0 /\partial \omega=0$ gives us
\begin{equation}
\me^{6\a+\lambda\phi} = \frac{3\kap \omega^2}{ V m_y }
\rfun{\sec^2}{\lambda\sqrt{\frac{3}{2\kap}}\a -\sqrt{\frac{3\kap}{2}}\phi},
\end{equation}
which is consistent with Eq.~\eqref{eq:classicalphantom}.

\subsection{WKB Gaussian wave packet}
\label{ssec:WKB-Gaussian}

The WKB Gaussian wave packet of quintessence models is expected to solve the
Wheeler--DeWitt equation in the WKB approximation. For the model with $m_x V<0$,
one
obtains
\begin{equation}
\psi^\text{WKB} = \rfun{C}{x,\omega} \me^{\frac{\mi}{\hslash} S_0},
\end{equation}
where
\begin{equation}
\rfun{C}{x,\omega} \coloneqq \frac{c}{\sqrt{\abs{\partial_x S_0}}}
=\frac{c}{\sqrt[4]{f/m_x^2}},\quad
f \coloneqq 6 \kap \omega^2 -2 m_x V \me^x,
\end{equation}
and two zeroth-order actions are
\begin{equation}
S_0=
\pm\frac{2\sqrt{f}}{ m_x}
	\mp \frac{ 2\omega\sqrt{6\kap} }{m_x}
	\rfun{\arccoth}{\omega\sqrt{\frac{6 \kap}{f}}},
\end{equation}
so that a Gaussian wave packet can be written as
\begin{equation}\label{eq:gaussianquinte}
 \varPsi=\int^{+\infty}_{-\infty} \dif
\omega\,\rfun{\mathcal{A}}{\omega,\bar\omega}
\me^{\frac{\mi}{\hslash}\tau\omega}\psi^\text{WKB},
\end{equation}
where $\mathcal{A}(\omega,\bar \omega)$ is the square root of a Gaussian
distribution
\begin{equation}\label{eq:gaussian}
\mathcal{A}(\omega,\bar \omega)=
\frac{1}{(\hslash\varGamma\sqrt{\pp})^{1/2}}\sfun{\exp}{-\frac{(\omega-\bar
\omega)^2}{2\hslash^2\varGamma^2}}.
\end{equation}
To integrate Eq.~\eqref{eq:gaussianquinte}, one can first expand $S_0$ around
$\bar \omega$,
\begin{equation}
S_0= \bar S_0 +( \partial_{\omega} \bar{S}_0) \Delta \omega
		+\frac{1}{2} (\partial_{\omega}^2 \bar{S}_0) \Delta \omega^2+o(\Delta
\omega^3),
\end{equation}
then apply the stationary phase approximation, and obtain the general form
of wave packet
\begin{equation}
\varPsi=\sqrt{\frac{\sqrt{\pp}}{\hslash\varGamma}}\rfun{C}{x,\bar\omega}
\me^{\frac{\mi}{\hslash}\a\bar \omega}
\frac{ \me^{P^2/4Q + \frac{\mi}{\hslash}\bar{S}_0}}{\sqrt{Q}}+\ldots
\end{equation}
The ellipsis denote the same formula but with the another $\bar S_0$, and
\begin{equation}
Q\coloneqq\frac{1}{2\varGamma^2 \hslash^2}
	-\frac{\mi}{2\hslash} \partial_{\omega}^2 \bar{S}_0,\quad
P\coloneqq\frac{\mi}{\hslash}(\tau+\partial_{\omega} \bar S_0).	
\end{equation}

For the phantom model, in the limit $\hbar\rightarrow 0$ the discreteness diappears and we can assume that the spectrum is continuous. Note that this approximation makes the leading order blind to the choice of the self-adjoint extension. The WKB wave packet is
$\psi^\text{WKB}=\rfun{C}{y,\omega}
\me^{\frac{\mi}{\hslash}S_0}$, with
\begin{equation}
S_0=\pm\frac{2\sqrt{f}}{ m_y}\mp
 \frac{ 2\omega\sqrt{6\kap} }{m_y}
\rfun{\arctan}{\frac{1}{\omega}\sqrt{\frac{f}{6\kap}}},
\end{equation}
and
\begin{equation}
\rfun{C}{y,\omega}\coloneqq\frac{c}{\sqrt[4]{f/m_y^2}},\quad
f\coloneqq-6\kap \omega^2 +2 m_y V \me^y.
\end{equation}

\subsection{Numerical matching}
\label{sec:num-matching}

The integral with Gaussian distribution in Eq.~\eqref{eq:gaussianquinte} cannot be
implemented analytically. Even though the WKB approximation
Sec.~\eqref{ssec:WKB-Gaussian} is effective, its precision is poor in regions where
semiclassical approach does not hold, for instance near the classical turning
point. Instead, one can turn to numerical approaches. 

\begin{figure}
\begin{subfigure}{.49\linewidth}
\includegraphics[width=\textwidth]{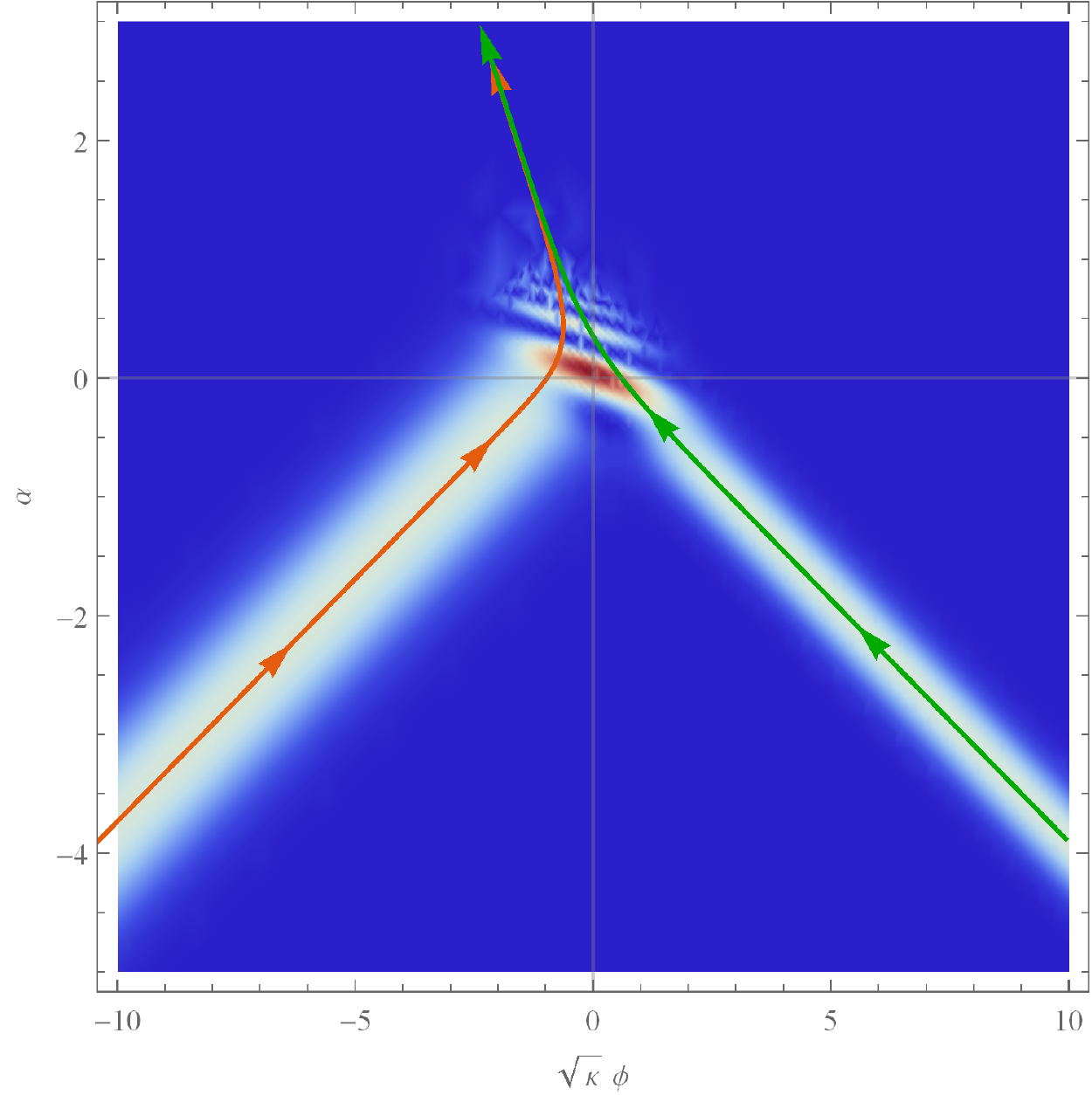}
\caption{$\BesselF_{\mi\nu}$, Schr\"odinger inner
product}
\label{fig:naive_sinh_F_2d}
\end{subfigure}
\begin{subfigure}{.49\linewidth}
\includegraphics[width=\textwidth]{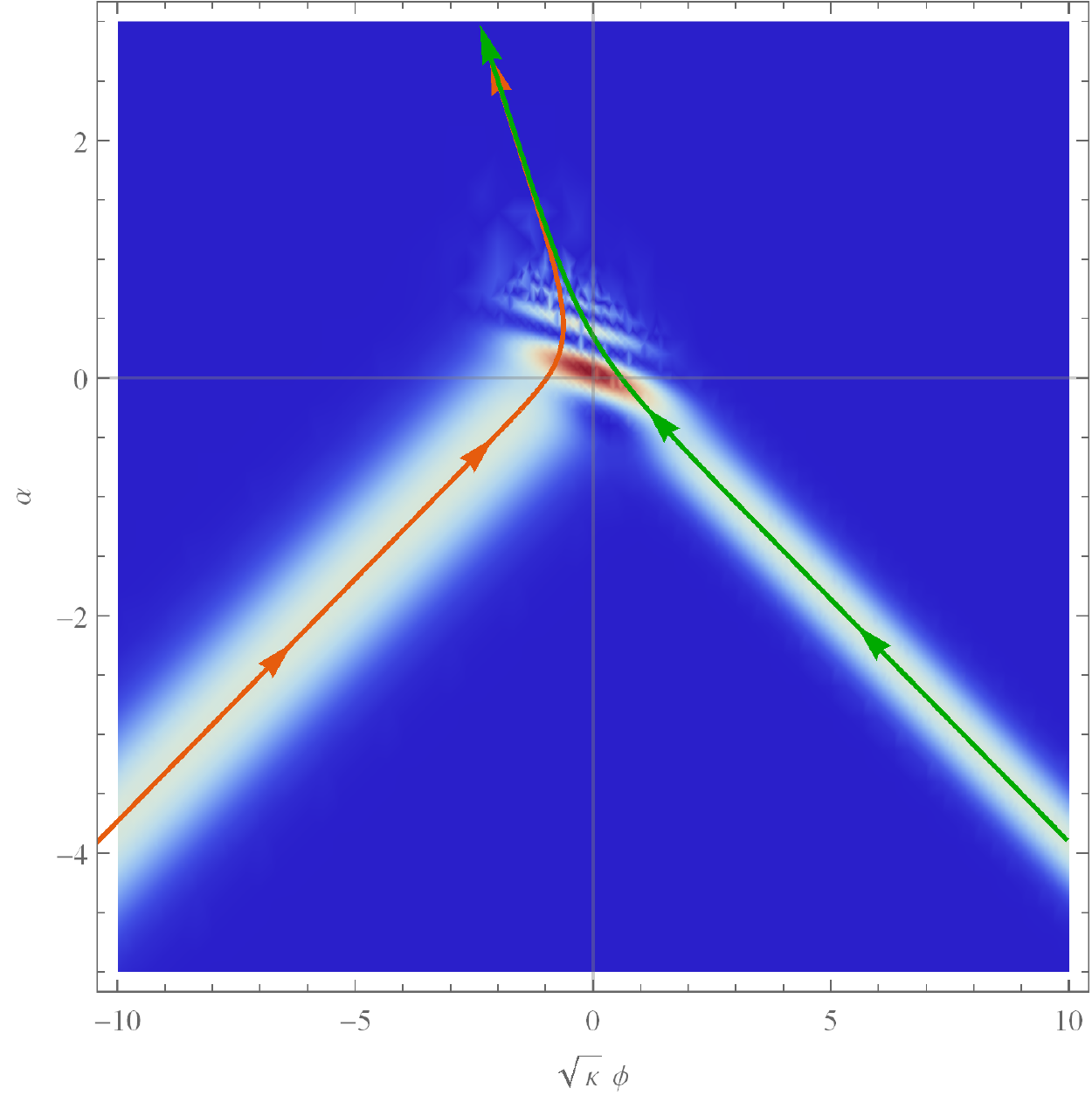}
\caption{$\BesselF_{\mi\nu}$, Mostafazadeh inner product}
\label{fig:mosta_sinh_F_2d}
\end{subfigure}

\begin{subfigure}{.49\linewidth}
\includegraphics[width=\textwidth]{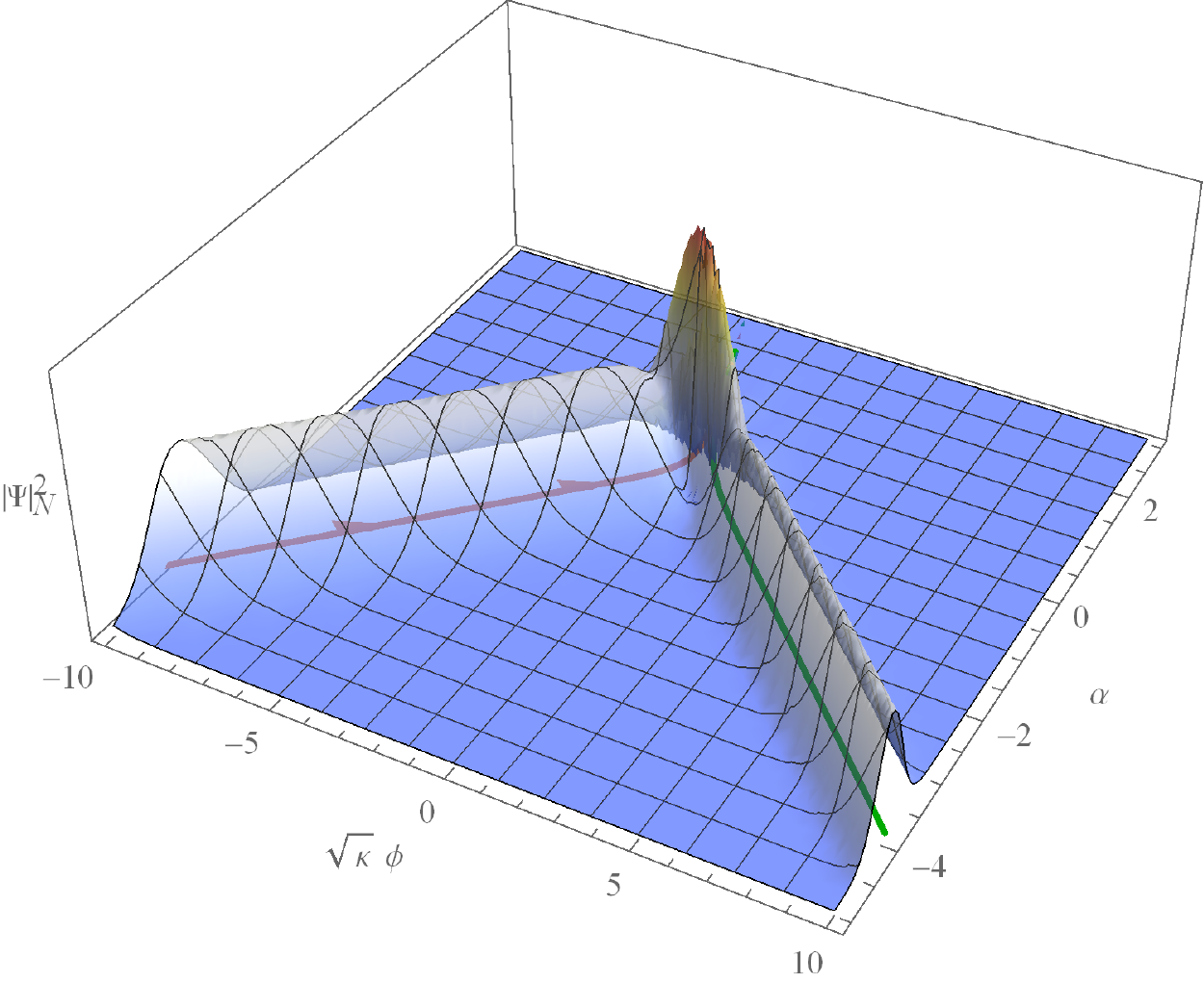}
\caption{$\BesselF_{\mi\nu}$, Schr\"odinger inner
product\label{fig:naive_sinh_F_3d}}
\end{subfigure}
\begin{subfigure}{.49\linewidth}
\includegraphics[width=\textwidth]{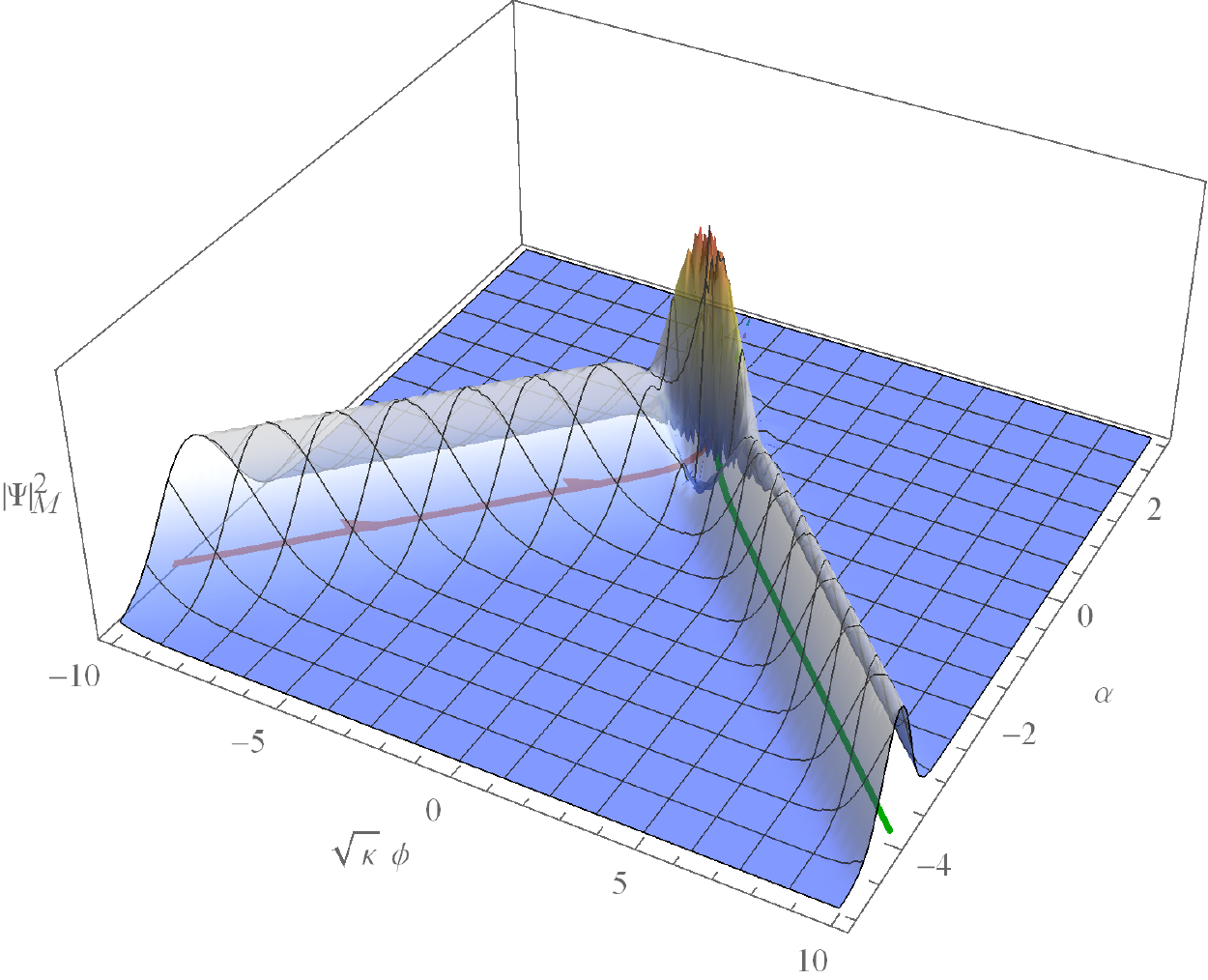}
\caption{$\BesselF_{\mi\nu}$, Mostafazadeh inner
product\label{fig:mosta_sinh_F_3d}}
\end{subfigure}
\caption{Wave packets and the corresponding classical trajectories for
quintessence with $m_x V<0$.
Parameters are $V = +1$, $\lambda = 4/5$, $\bar\omega = -35/8$ and
$\varGamma = 7/5$.
\label{fig:sinh-F}}
\end{figure}

\begin{figure}
\begin{subfigure}{.49\linewidth}
\includegraphics[width=\textwidth]{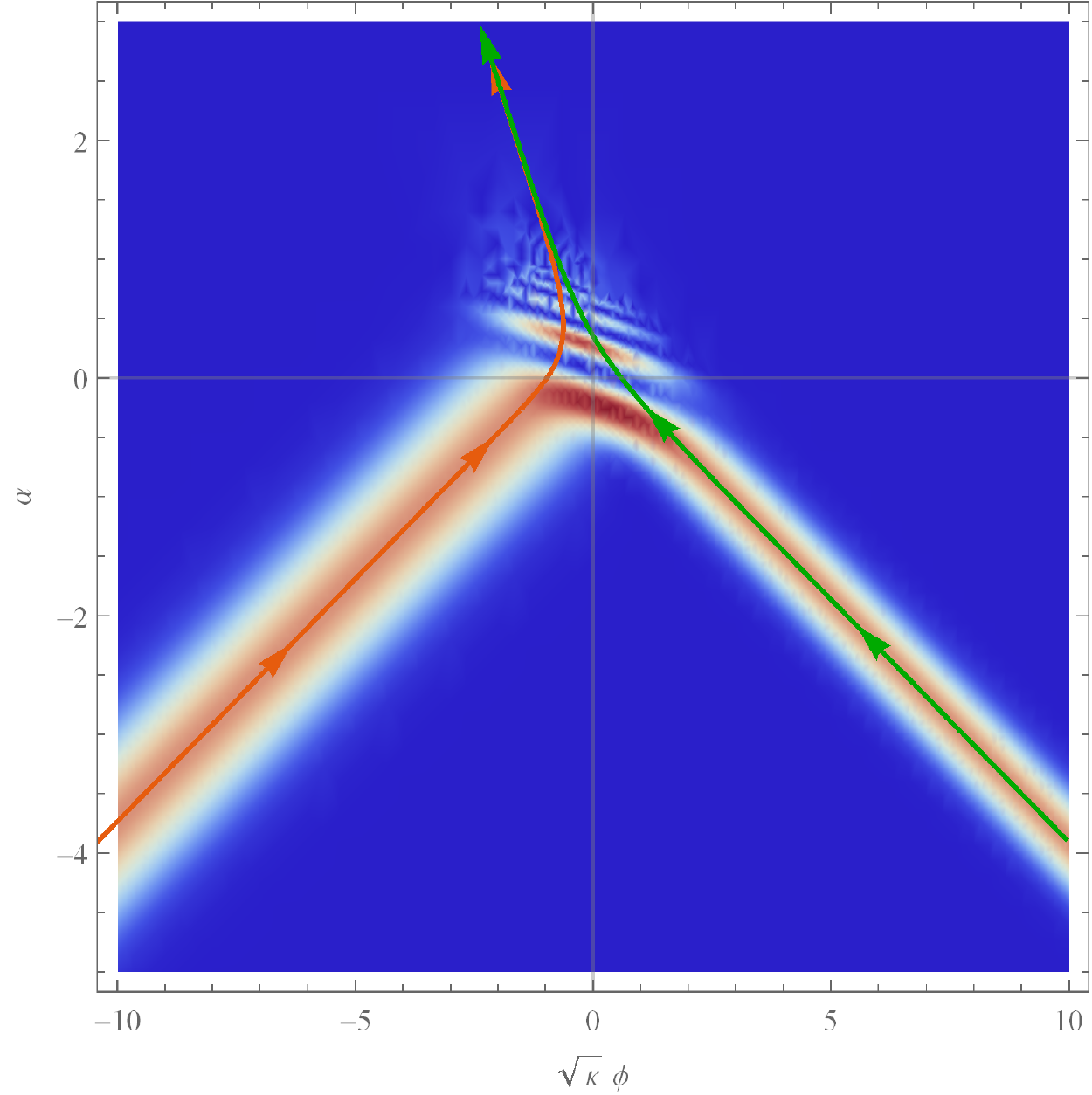}
\caption{$\BesselG_{\mi\nu}$, Schr\"odinger inner
product}
\label{fig:naive_sinh_G_2d}
\end{subfigure}
\begin{subfigure}{.49\linewidth}
\includegraphics[width=\textwidth]{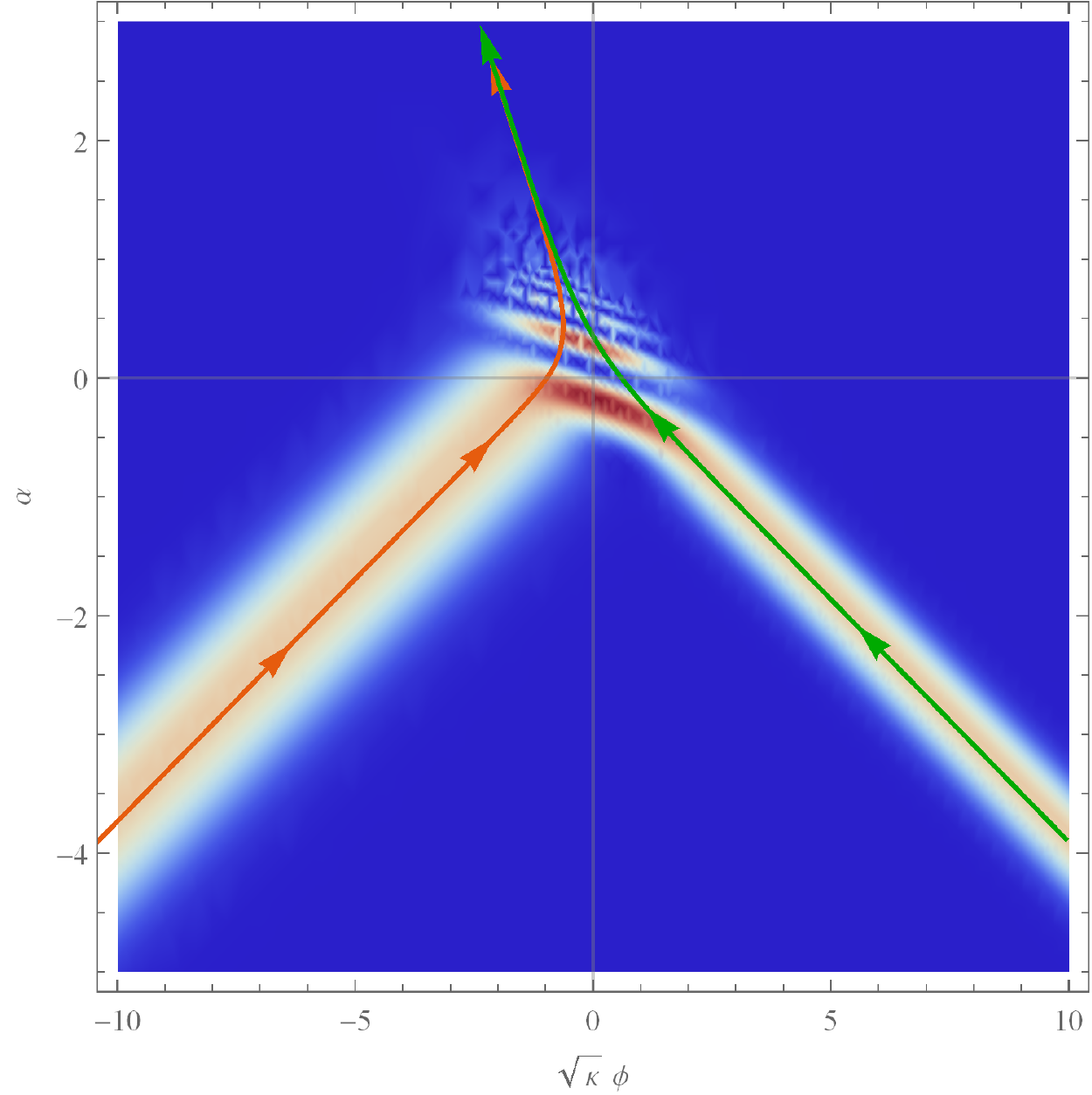}
\caption{$\BesselG_{\mi\nu}$, Mostafazadeh inner
product}
\label{fig:mosta_sinh_G_2d}
\end{subfigure}

\begin{subfigure}{.49\linewidth}
\includegraphics[width=\textwidth]{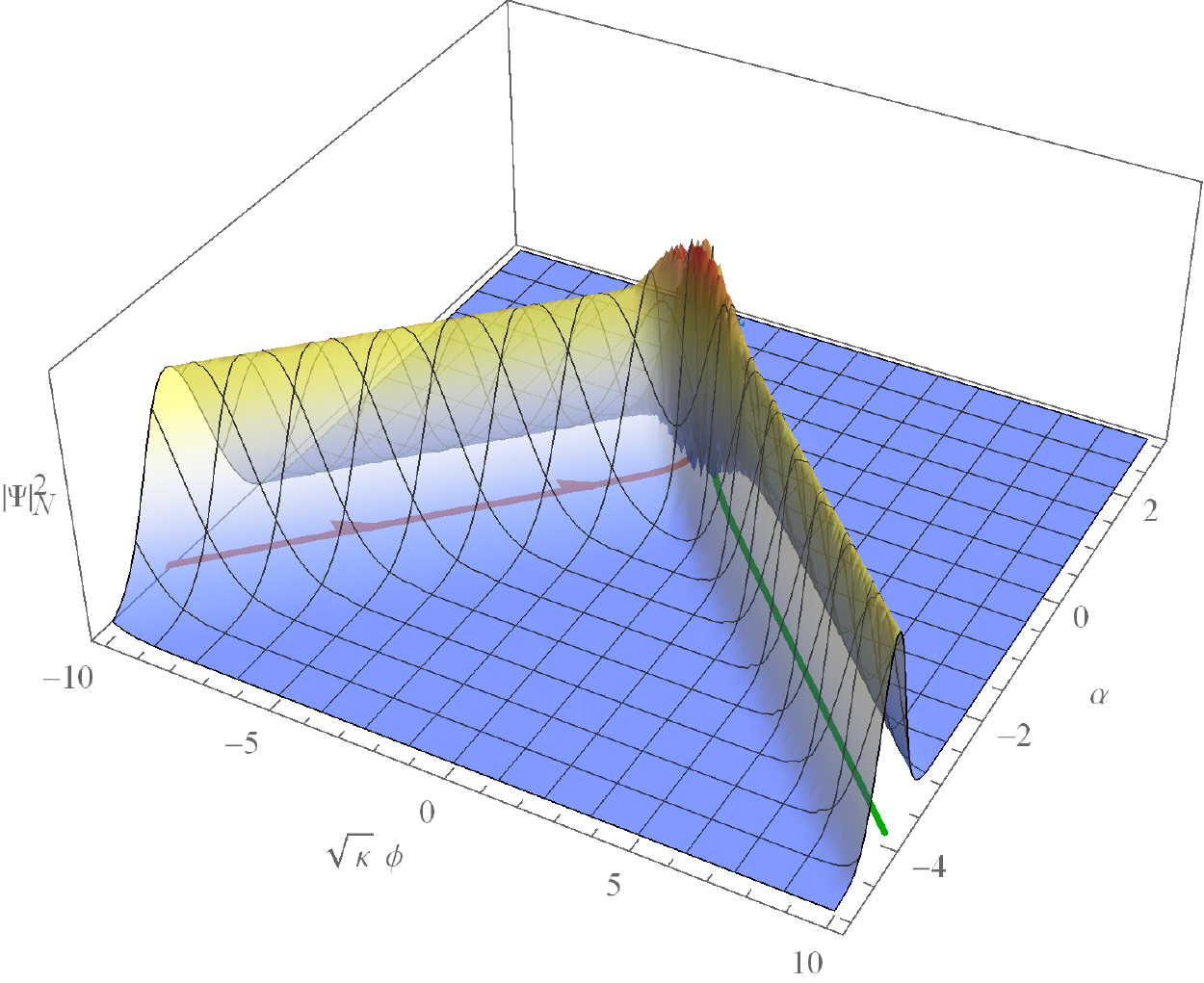}
\caption{$\BesselG_{\mi\nu}$, Schr\"odinger inner
product\label{fig:naive_sinh_G_3d}}
\end{subfigure}
\begin{subfigure}{.49\linewidth}
\includegraphics[width=\textwidth]{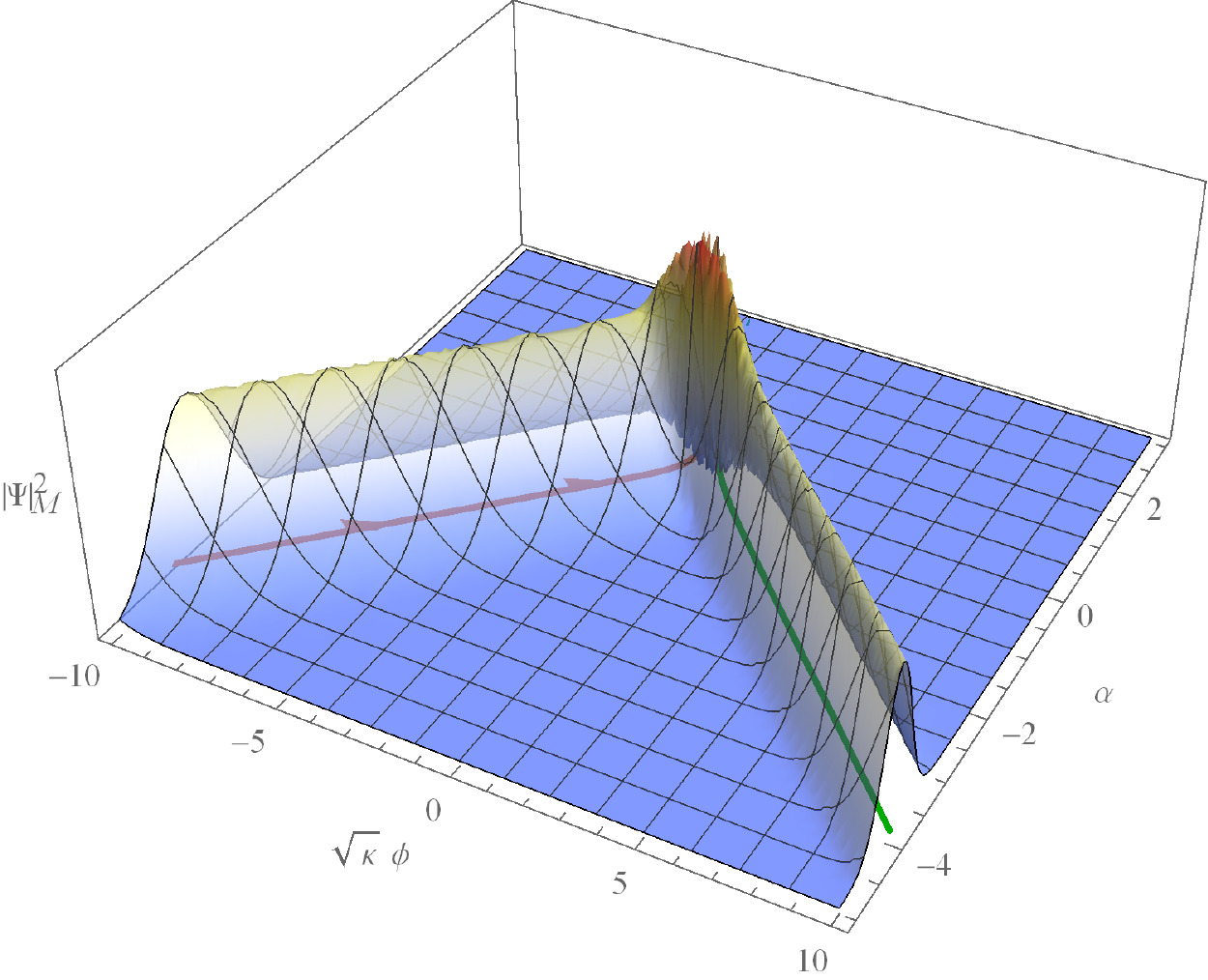}
\caption{$\BesselG_{\mi\nu}$, Mostafazadeh inner
product}
\label{fig:mosta_sinh_G_3d}
\end{subfigure}
\caption{Wave packets and the corresponding classical trajectories for the
quintessence with $m_x V<0$.
Parameters are $V = +1$, $\lambda = 4/5$,
$\bar\omega = -35/8$ and $\varGamma = 7/5$.
}\label{fig:sinh-G}
\end{figure}

\begin{figure}
\begin{subfigure}{.49\linewidth}
\includegraphics[width=\textwidth]{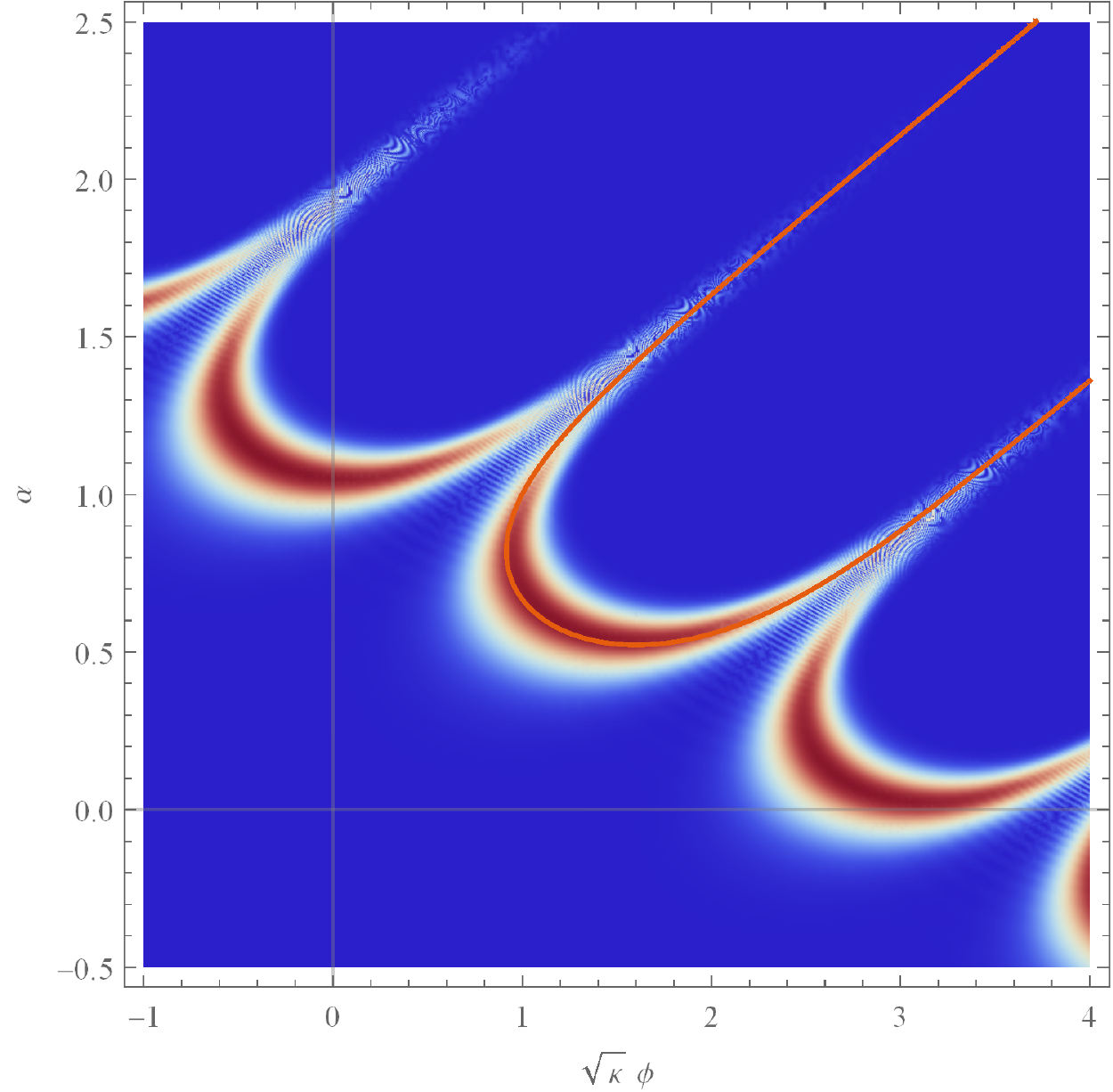}
\caption{Schr\"odinger inner product}
\label{fig:naive_cos_1_2d}
\end{subfigure}
\begin{subfigure}{.49\linewidth}
\includegraphics[width=\textwidth]{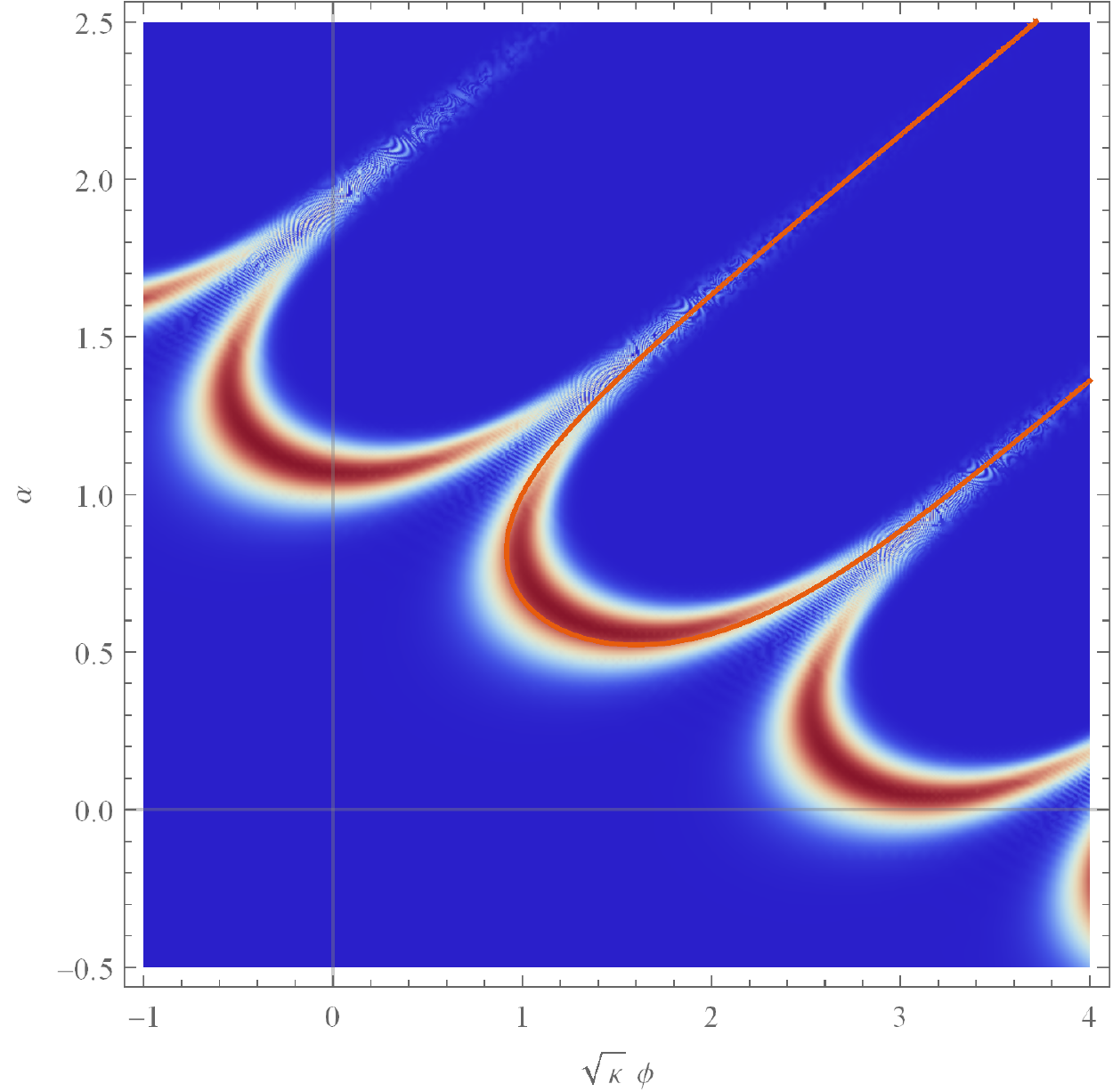}
\caption{Mostafazadeh inner product}
\label{fig:mosta_cos_1_2d}
\end{subfigure}

\begin{subfigure}{.49\linewidth}
\includegraphics[width=\textwidth]{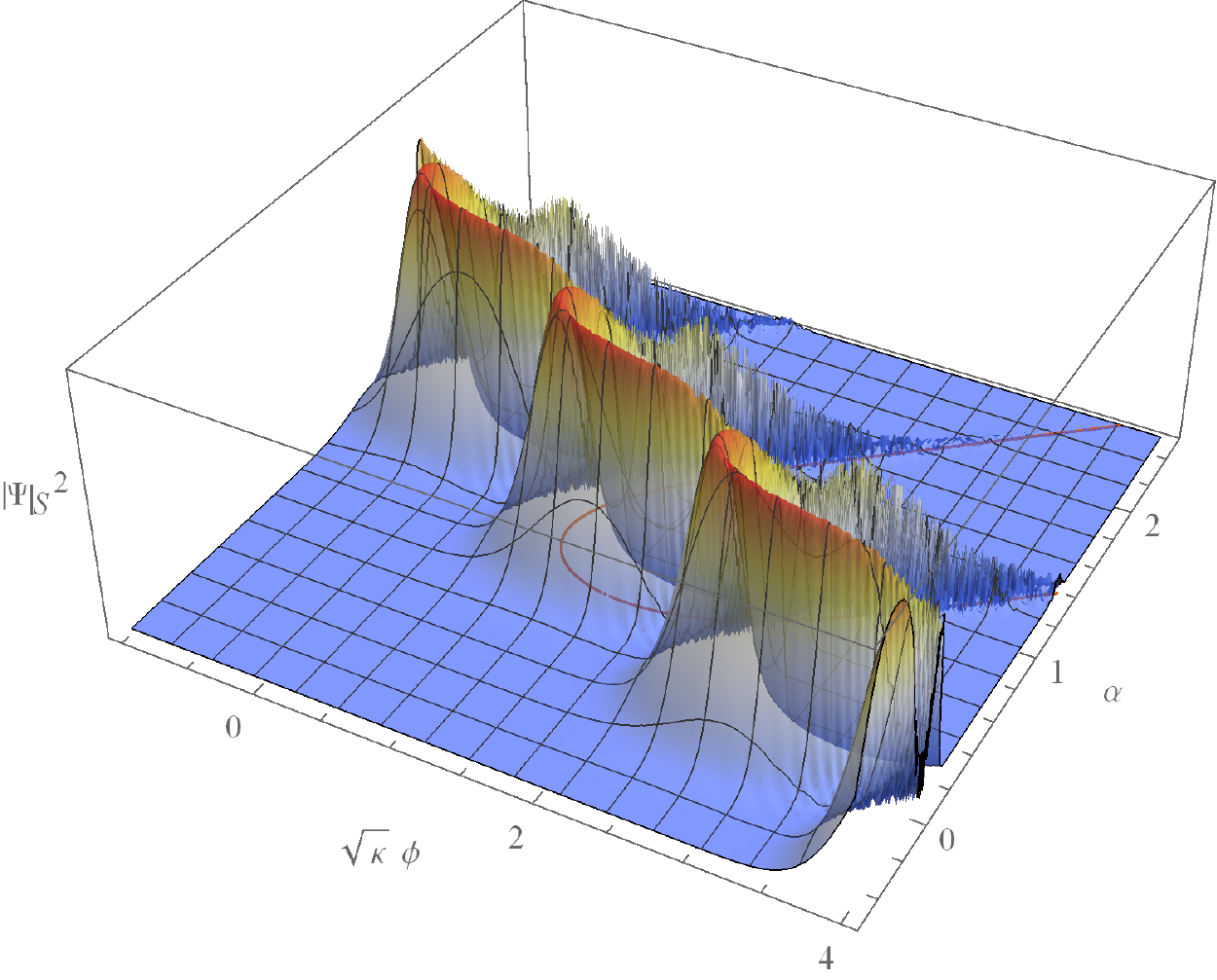}
\caption{Schr\"odinger inner product}
\label{fig:naive_cos_1_3d}
\end{subfigure}
\begin{subfigure}{.49\linewidth}
\includegraphics[width=\textwidth]{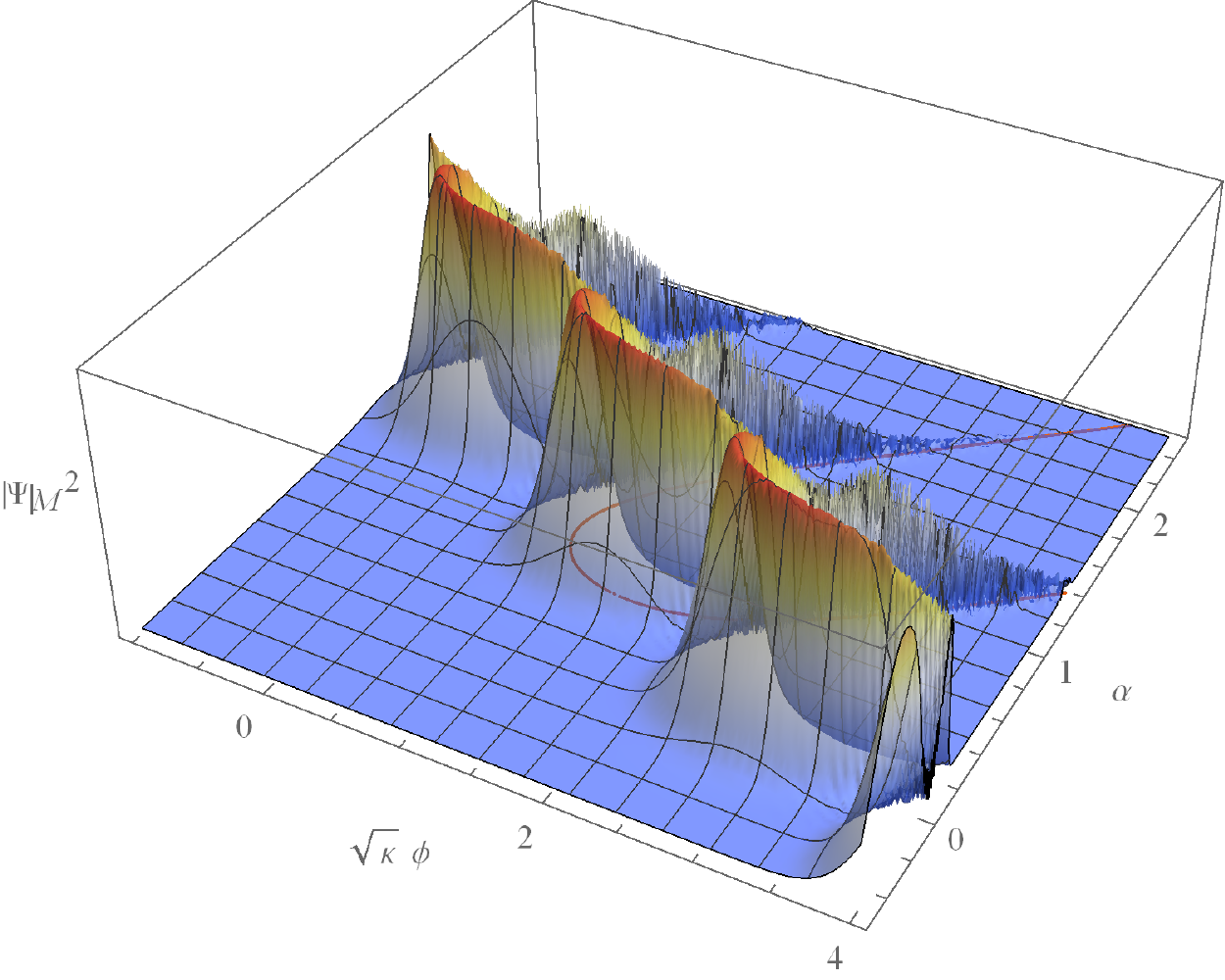}
\caption{Mostafazadeh inner product}
\label{fig:mosta_cos_1_3d}
\end{subfigure}
\caption{Poissonian wave packets established by the wave functions with
discrete spectrum and the corresponding classical trajectories for the phantom
model. Parameters are $V = 1$, $\lambda = 2$, $\bar\omega^2 = 10240/9$,
meanwhile $a=1$ is fixed.
}\label{fig:cos-poisson}
\end{figure}

With the wave functions normalized, one may construct wave packets for the
quintessence and phantom models. The corresponding plots are in
\eqref{fig:sinh-F}, \eqref{fig:sinh-G} and \eqref{fig:cos-poisson}. 
The parameters are specified in Planck units $\hbar=\kap=1$. 
The common feature of the plots is that the
wave
packets coincide with classical trajectories and follow them as closely as
possible. The height of the wave `tube' is negatively correlated to the `speed'
of the classical trajectory with respect to the Klein--Gordon time $\tau$,
i.e.\ the higher the `speed' is, the lower the amplitude of the wave `tube' is
\cite{hawking1986}. It is interesting to note that for all models
the naive inner product Eq.~\eqref{eq:naive-L2-norm} happen to approximate
the conserved norm Eq.~\eqref{eq:KG-Mos} very well so that there's no noticeable
difference in plots.

In Figs.~\eqref{fig:sinh-F} and \eqref{fig:sinh-G}, the classical trajectory contains two disjoint
branches representing two distinct solutions separated by cosmological
singularity. This leads to a quite interesting interference of the two wave tubes. The different choice of $a$ corresponds to slightly different wave packets.

For the phantom model, on the other hand, Poisson's distribution of
momentum (see Fig.~\eqref{fig:cos-poisson}) has been chosen,
\begin{equation}
\mathcal{A}_n = \frac{\bar n^{n/2}
\me^{-\bar n/2}}{\sqrt{n!}}.
\end{equation}
As expected from Eq.~\eqref{eq:periodicity} the wave packet is periodic in $\tau$
emerging along all the periodic classical solutions separated by Big Rip singularities.


\section{Conclusions}
\label{sec:conclution}

In this paper, by using the integral of motion to eliminate the lapse
function in Friedmann equation, we have solved the cosmological model with
Liouville field for homogeneous isotropic metrics.
The general classical solutions are obtained and represented in terms of minisuperspace
variables only, such that the correspondence between classical and
quantum theory can be demonstrated manifestly.
The quantum wave packets reproduce the classical limit in a sense that
the distributions of traditional Schr\"odinger's norm and the Mostafazadeh's inner product 
are maximized near the classical trajectories.

The classical models of quintessence with potential unbounded below
and the phantom fields give rise to the appearance of a family of non-equivalent
quantum models, because the energy density operators are not essentially self-adjoint
operator. In order to preserve unitarity and correct classical limit one has to omit half of the spectrum. While this
requires that the wave packet at some fixed $\tau$ belongs to much narrower class
than $\rfun{L^2}{\mathbb{R}}$, it is enough to produce wave packets in the vicinity of the classical
trajectories.

For the phantom field the resulting spectrum is discrete. It is associated
with the fact that at the classical level the universe exists in a finite interval between
two singularities and non-singular unitary evolution is accessible through the periodicity
of wave function. This periodicity may be regarded as a fundamental condition not only for the homogeneous
but also on inhomogeneous modes. On the other hand, if
 the minisuperspace wave packet contains multiple semiclassical branches they may be associated with coherent 
superposition of different universes. This Schr\"odinger-cat-like
effect at the cosmic scale might be an artefact of the model in minisuperspace.
In the full theory in Wheeler's superspace \cite{Wheeler1968a}, inhomogeneity is
involved, which may serve as an unobservable environment, in contrast with the
scale factor \cite{Kiefer2000a}. The observable effects are then fully described
by the density matrix of the scale factor only, whose off-diagonal elements
characterize the superposition of universes with different scale factors.
Calculation suggests that those elements are highly-suppressed in the
above-mentioned decoherence scheme \cite{kiefer1987,Kiefer1992a}; hence the
cosmic Schr\"odinger cat might be fictitious, and the superposition of distinct
semiclassical branches might be decohered to vanish.
The approach developed in the paper can also be extended to Higher dimensional \cite{garcia2007,letelier2010} and anisotropic models, such as Bianchi-\rom{1} cosmology considered in \cite{kamenshchik2017a}.

As the different self-adjoint extensions lead to different quantum evolution and require the wavefunction to belong to the different restricted functional class they may produce different observable results. The leading order of WKB approximation is insensitive however one may expect that the choice of self-adjoint extension should be improtant for the NLO corrections to the spectra of perturbations \cite{KK1,Bini:2013fea,BKK1,BKK2}.


\providecommand{\href}[2]{#2}\begingroup\raggedright\endgroup

\end{document}